\documentclass[fleqn,usenatbib]{mnras}
\usepackage{natbib}
\usepackage{newtxtext,newtxmath}

\usepackage[T1]{fontenc}

\DeclareRobustCommand{\VAN}[3]{#2}
\let\VANthebibliography\thebibliography
\def\thebibliography{\DeclareRobustCommand{\VAN}[3]{##3}\VANthebibliography}

\usepackage{graphicx}	
\usepackage{amsmath}	
\usepackage{xcolor}
\usepackage{gensymb}
\usepackage{multirow}
\newcommand{\edit}{}

\newcommand{\rev}{}

\title[Warping and breaking protoplanetary discs]{On the conditions for warping and breaking protoplanetary discs}

\author[A. K. Young et al.]{
Alison K. Young,$^{1,2}$\thanks{E-mail: alison.young@ed.ac.uk (AKY)}
Struan Stevenson,$^{1,2}$
C.~J. Nixon,$^{3}$
and Ken Rice$^{1,2}$
\\
$^{1}$SUPA, Institute for Astronomy, University of Edinburgh, Blackford Hill, Edinburgh EH9 3HJ, UK\\
$^{2}$Centre for Exoplanet Science, University of Edinburgh, Edinburgh, EH9 3HJ, UK\\
$^{3}$School of Physics and Astronomy, Sir William Henry Bragg Building, Woodhouse Ln., University of Leeds, Leeds LS2 9JT, UK\\
}

\date{Accepted XXX. Received YYY; in original form ZZZ}

\pubyear{2023}

\begin{document}
\label{firstpage}
\pagerange{\pageref{firstpage}--\pageref{lastpage}}
\maketitle

\begin{abstract}
Recent observations demonstrate that misalignments and other out-of-plane structures are common in protoplanetary discs. Many of these have been linked to a central host binary with an orbit that is inclined with respect to the disc. We present simulations of misaligned circumbinary discs with a range of parameters to gain a better understanding of the link between those parameters and the disc morphology in the wave-like regime of warp propagation that is appropriate to protoplanetary discs. The simulations confirm that disc tearing is possible in protoplanetary discs as long as the mass ratio, $\mu$, and disc-binary inclination angle, $i$, are not too small. For the simulations presented here {\edit this corresponds to $\mu > 0.1$ and} $i \gtrsim 40^\circ$. For highly eccentric binaries, tearing can occur for discs with smaller misalignment. Existing theoretical predictions provide an estimate of the radial extent of the disc in which we can expect breaking to occur. However, there does not seem to be a simple relationship between the disc properties and the radius within the circumbinary disc at which the breaks appear, and furthermore the radius at which the disc breaks can change as a function of time in each case. We discuss the implications of our results for interpreting observations and suggest some considerations for modelling misaligned discs in the future.
\end{abstract}

\begin{keywords}
 accretion, accretion discs -- hydrodynamics -- protoplanetary discs
\end{keywords}



\section{Introduction}

High-resolution images reveal many kinds of substructures within protoplanetary discs. These include out-of-plane structures such as warps and misaligned inner disc regions that are inferred from kinematics \citep[e.g.][]{rosenfeld2012,perez2018,bi2020,paneque2021} and/or the shadows they cast across the outer disc regions \citep[e.g.][]{marino2015,benisty2017,muro-arena2020,laws2020}. Some young protostars display variability that can be explained by the obscuration of the star by out-of-plane structures or variable accretion rates \citep[e.g.][]{fang2014,ansdell2020,lakeland2022}, both of which may be due to disc warps. Circumbinary discs around long period and/or eccentric binaries ($P>30$~days, $e>0.2$) are observed to have mutual inclinations ranging from co-planar to polar \citep{czekala2019} and circumstellar discs in binaries tend not to be coplanar \citep{rota2022}. It is clear that misaligned discs are common and that the interactions between discs and misaligned host binaries will play an important role in disc evolution and planet formation in these systems.

Now that we are able to observe protoplanetary disc warps and misalignments, the next step is to link these structures to the mechanisms that are driving them. This requires a two-pronged approach of both understanding which properties can be derived from observations of misaligned discs and elucidating the links between the fundamental disc properties (inclination, semithickness, viscosity, density profile etc.) and the resulting structure. The former problem has been addressed by \citet{juhasz2017,min2017,facchini2018,young2021} and \citet{young2022}. Most of this work indicates that only moderate to large misalignments produce detectable structures. (There are some observations of azimuthal asymmetries in scattered light however that might be due to low amplitude warps; e.g. \citealt{debes2017}.) In this paper, we turn to the latter question by simulating the evolution of a selection of misaligned circumbinary discs with different properties. 


Circumbinary accretion discs become warped when the orbital plane of the disc is inclined with respect to that of the central binary. The misalignment generates a torque that causes the orbits of the disc gas to precess. The torque acts in the direction defined by the cross product of the binary and disc angular momentum vectors, and the magnitude is given by
\begin{equation}
\label{eq:Gp}
    G_{\rm p} = \frac{3}{4}\eta{a^2} \Sigma\Omega^2\cos{i}\sin{i},
\end{equation}
where $\eta=m_1m_2/(m_1+m_2)^2$ is the reduced mass of the binary, $\Sigma$ is the surface density, $\Omega$ is the orbital frequency and $i$ is the binary-disc misalignment \citep[see e.g.][]{nixon2013,facchini2013}. It follows that the precession torque decreases with radius and the outcome of this differential precession is a warped shape. If the warp in the disc becomes too strong then it can break \citep{dogan2018}.

The mechanism by which the warp is communicated through the disc depends on the \cite{shakura1973} disc viscosity parameter $\alpha_{\rm SS}$, and the disc angular semithickness $H/R$ \citep{papaloizou1983}. When $\alpha_{\rm SS} > H/R$ the warp propagation follows a diffusion equation; this is expected to occur in thin, fully-ionised discs (e.g. in the high accretion rate states of X-ray binaries and AGN where the disc orbits a black hole). In protoplanetary discs, the effective viscosity is expected to be significantly lower and typically $\alpha_{\rm SS} < H/R$. In this regime, the warp propagates via bending waves \citep{papaloizou1995,lubow2000} as long as the disc is nearly Keplerian, which requires \citep{ogilvie1999}
\begin{equation}
\label{eq:kepcondition}
    \left|\frac{\Omega^2 -\kappa^2}{\Omega^2}\right|\lesssim H/R\,,
\end{equation}
where, $\kappa$ is the epicyclic frequency given by
\begin{equation}
    \kappa^2 = 4\Omega^2 + 2R\Omega \frac{d\Omega}{dR}\,.
\end{equation}

Early simulations of misaligned circumstellar discs in binaries and circumbinary discs demonstrated that the disc becomes warped and also that a large misalignment can cause the disc to separate into two components \citep[][see also \citealt{fragner2010}]{larwood1996,larwood1997}. These also highlighted the dependence of the structure on the communication within the disc, equivalent to the disc thickness: the disc was more distorted with a higher Mach number (thinner disc). Later works reaffirmed these effects, showing that the shape of the warped disc depends on the viscosity and the amplitude of the external torque \citep[e.g.][]{facchini2014}. \citet{foucart2014} shed some doubt on the possibility of observing warped protoplanetary (circumbinary) discs as their estimate of the alignment timescale was significantly shorter than the lifetime of the disc. However, this estimate assumed $\alpha=0.01$ but this is likely to be an order of magnitude smaller \citep[e.g.][]{rosotti2023}, similarly increasing the alignment timescale.\footnote{It is well-established that in fully-ionised discs the viscosity is vigorous with typical measurements of $\alpha \approx 0.2-0.4$ \citep[e.g.][]{martin2019}. However, in discs where the material is not fully ionised the effects of MHD turbulence can be sharply reduced \citep{gammie1998}. The historical estimate of $\alpha \sim 0.01$ in protoplanetary discs, which comes from e.g. modelling of the evolution of disc radii \citep{hartmann1998}, represents a time averaging of these high (fully ionised, strong turbulence) and low (low ionisation, weak turbulence) viscosity states.}

When the disc becomes strongly warped, the warp can steepen into a break where two regions of the disc occupy distinctly different planes and are joined by a low-density region where the disc plane changes abruptly \citep{larwood1997,fragner2010,nixon2012apr,nixon2012}. Using the analysis of \cite{ogilvie1999,ogilvie2000} appropriate for the diffusive regime of warp propagation, \cite{dogan2018} show that this behaviour can be explained as an instability in the propagation of the disc warp, and this is confirmed by numerical simulations \citep{raj2021a}. {\edit\cite{dogan2020} extend the analysis of \cite{dogan2018} to include non-Keplerian discs.} \cite{drewes2021} argue that the breaking that occurs in wave-like discs is fundamentally the same instability due to additional dissipation present in a strong warp that locally enhances the effective viscosity such that the disc locally behaves in a diffusive manner. In the context of black hole accretion discs, simulations show that {\edit both diffusive and wave-like discs} can be torn into multiple disconnected rings either due to Lense-Thirring precession or due to a central binary {\edit \citep{nixon2012,nixon2013,nealon2015,liska2021,drewes2021}}. In the wave-like regime, simulations {\edit find that a protoplanetary disc may break into just two components or form a single misaligned inner ring \citep[e.g.][]{facchini2018,hirsh2020}}. Such a ring has been observed in a circumtriple protoplanetary disc and its origin has been linked to the misalignment between the disc and stellar orbits \citep{kraus2020}.

Where the disc breaks, or where the warp amplitude is maximised, is an interesting question. It determines the detectability of these structures and helps address the question of what is triggering an observed warp or break.
The modelling of warped protoplanetary disc evolution has been hampered by the requirement for wave-like warp propagation. The smoothed particle hydrodynamics (SPH) method is most commonly used for such simulations, and this method includes explicit artificial viscosity which can, at low resolution, result in a large effective viscosity. This means that sufficient resolution is required to accurately model a wave-like disc \citep[see the discussion in Section 2.4 of][]{drewes2021}. Previous studies of misaligned circumbinary discs have concentrated on verifying linear analytical theory with hydrodynamical simulations \citep[e.g.][]{facchini2013} or a limited range of parameters \citep[e.g.][]{facchini2014,foucart2014}. {\edit Here we want to explore the range of conditions for which the disc breaks and for which the disc merely warps to assist with the interpretation of observations.} We will draw comparisons with prior work too where possible. 

The purpose of this paper is to determine the degree of warping of a circumbinary disc for a range of parameters and to advance our understanding of the conditions under which the disc is likely to tear. In the following section we describe the simulation setup and in Section \ref{sec:results} we present the results of simulations testing each parameter in turn. Sections \ref{sec:discussion_factors} and \ref{sec:implications} discuss the structures produced for different disc and binary properties and the implications for interpreting observations. In section \ref{sec:lessons} we make suggestions for future modelling of misaligned and warped protoplanetary discs. 

\section{Method}

\subsection{Hydrodynamical model}
\label{sec:hydro}
We use the publicly available SPH code {\sc phantom} \citep{price2018aa} to perform the simulations in this paper. {\sc phantom} has been used extensively to model warped discs and was benchmarked against the nonlinear theory of \cite{ogilvie1999} by \cite{lodato2010}. {\sc phantom} has also been used extensively to model circumbinary discs \citep[starting with][]{nixon2012jul}. Ensuring the total viscosity remains low is a key issue when using SPH to model discs in the wave-like regime. This is because the artificial viscosity, with linear and quadratic coefficients $\alpha_{\rm SPH}$ and $\beta_{\rm SPH}$ respectively \citep[see][for details]{price2018aa}, is resolution dependent, meaning that level of numerical viscosity present in the disc can be too large if insufficient resolution is employed \cite[see, e.g.,][]{meru2012}. {\edit Following \cite{fragner2010}, who used a grid-based code, \citet{nealon2015} showed using {\sc phantom} that an imposed warp in a Keplerian potential can propagate in a wave-like fashion. \citet{drewes2021} demonstrated that with sufficiently high resolution the (numerical) viscosity remains sufficiently low that reasonable agreement can be reached between the results of {\sc phantom} SPH simulations in the linear regime and the linearised wave-like warp evolution equations of \cite{lubow2000}. \cite{drewes2021} also showed that the disc tearing behaviour is adequately converged in terms of the disc shape and density profile at the point the disc breaks (see their Figs 5 \& 6).} In the simulations presented here we employ the default artificial viscosity in {\sc phantom} which comprises setting $\beta_{\rm SPH}=2$ and allowing $\alpha_{\rm SPH}$ to vary between 0 and 1 using the \citet{cullen2010} switch. In practice, $0.05<\alpha_{\rm SPH}<0.2$ in most of the disc. No additional physical viscosity is imposed.

\subsection{Simulation parameters}
\label{sec:sim_params}
We perform a range of simulations of circumbinary discs varying several system parameters. The central binary is modelled with a pair of sink particles \citep{bate1995aa} in an initially circular orbit of binary separation $a_{\rm{B}} = 1$~au. The total mass of the stellar binary is 2~M$_\odot$ and the mass of each component is chosen to give a binary mass ratio of  $\mu=m_2/(m_1+m_2)=\{0.1,0.2,0.3,0.5\}$. The sink particles are free to accrete gas particles from the disc and have accretion radii of 0.2~au. The aspect ratio $H/R=0.05$ at $R=2.1$~au. The surface density profile is set via $\Sigma(R)= \Sigma_0 (R/2.1{\rm au})^{-p}$ where $\Sigma_0=477$~g~cm$^{-2}$ and $p=0.5$ unless otherwise specified, giving $m_{\rm disc}=0.01$~M$_\odot$ for $R_{\rm out}=50$~au. An isothermal equation of state is implemented with the sound speed profile given by $c_{\rm s}(R) = c_{{\rm s},0}(R/2.1{\rm au})^{-q}$, with $q=0.25$ unless otherwise specified.

We investigate the effect on the disc evolution of: the binary mass ratio $\mu$, the initial inclination of the disc with respect to the binary $i$, the sound speed profile exponent $q$, the density profile exponent $p$, the disc outer radius $R_{\rm out}$, and the binary eccentricity $e$. The details of these are given in the relevant results sections {\edit and a summary of the parameters for each simulation is provided in Table \ref{tab:sim_parameters}}.
The initial inner radius $R_{\rm in}$ is set from estimates of the tidal truncation radius from \citet{miranda2015} and the values are displayed in Table~\ref{tab:sim_parameters}. For more details see section \ref{sec:inneredge}. The circumbinary disc is modelled with $10^7$ SPH particles for $R_{\rm out} = 50$~au.

\begin{table*}
    \centering
    \begin{tabular}{p{2cm} l l l l l l l}
       & $\mu$ & $e$ & $i$ [deg]  & $p$ & $q$ & $R_{\rm {in}}$ [au] & $R_{\rm {out}}$ [au] \\
        \hline
        \multirow{3}{2cm}{Binary mass ratio} & 0.1  & 0 & 45 & 0.5 & 0.25 & 1.3 & 50 \\ 
         & 0.2 & 0 & 45 & 0.5 & 0.25 & 2.1 & 50 \\ 
         & 0.5 & 0 & 40 & 0.5 & 0.25 & 2.1 & 50 \\ 
         \hline
         \multirow{5}{2cm}{Misalignment} & 0.3 & 0 & 30 & 0.5 & 0.25 & 2.1 & 50 \\
         & 0.3 & 0 & 40 & 0.5 & 0.25 & 2.1 & 50 \\
         & 0.3 & 0 & 60 & 0.5 & 0.25 & 1.3 & 50 \\
         & 0.3 & 0 & 120 & 0.5 & 0.25 & 1.3 & 50 \\
         & 0.3 & 0 & 150 & 0.5 & 0.25 & 1.3 & 50 \\
         \hline
         \multirow{2}{2cm}{Sound speed profile} & 0.3 & 0 & 30 & 0.5 & 0.25 & 2.1 & 50 * \\ 
         & 0.3 & 0 & 30 & 0.5 & 0.5 & 2.1 & 50 \\ 
         \hline
         \multirow{2}{2cm}{Surface density profile} & 0.3 & 0 & 40 & 0.5 & 0.25 & 2.1 & 50 *\\ 
         & 0.3 & 0 & 40 & 1.5 & 0.25 & 2.1 & 50 \\ 
         \hline
          \multirow{2}{2cm}{Disc outer radius} & 0.3 & 0 & 30 & 0.5 & 0.25 & 2.1 & 25 \\ 
         & 0.3 & 0 & 30 & 0.5 & 0.25 & 2.1 & 40 \\ 
         & 0.3 & 0 & 30 & 0.5 & 0.25 & 2.1 & 50 *\\
         \hline
         \multirow{2}{2cm}{Binary eccentricity} & 0.5 & 0.4 & 30 & 0.5 & 0.25 & 2.1 & 50 \\ 
         & 0.5 & 0.6 & 30 & 0.5 & 0.25 & 2.1 & 50 \\ 
         & 0.5 & 0.8 & 30 & 0.5 & 0.25 & 2.1 & 50 \\
         \hline
         \multirow{2}{2cm}{Disc inner radius} & 0.3 & 0 & 60 & 0.5 & 0.25 & 1.3 & 50 * \\ 
         & 0.3 & 0 & 60 & 0.5 & 0.25 & 2.2 & 50 \\
    \end{tabular}
    \caption{\edit Summary of the parameters used in the simulations presented in this paper. The symbols are defined in full in section \ref{sec:sim_params}. The initial values of the inner radius are set to the locations of the outer Lindblad resonances likely to be truncating the disc (see section \ref{sec:inneredge} for further discussion). *Duplicate of simulation already listed.}
    \label{tab:sim_parameters}
\end{table*}

\section{Results}
\label{sec:results}
In this section we present the results of simulations to assess how the physical properties of the disc and host binary affect its structure. To analyse the simulated discs we consider {\edit concentric spherical shells within the disc model} to examine the evolution of the radial density profile, orientation and warp amplitude. {\edit We use spherical polar coordinates to account for the different inclinations of the various discs \citep{petterson1977}.} The tilt $\beta(R,t)$, describes the misalignment of a ring of radius $R$ within the disc with respect to the total angular momentum vector (binary + disc). The twist $\gamma(R,t)$ refers to the transformation in the azimuthal direction (sometimes called phase). The warp amplitude is a dimensionless measure of the distortion of the disc:
\begin{equation}
    \psi=R\left|\frac{\partial \hat{l}}{\partial R}\right|,
\end{equation}
where $\hat{l}(R,t)$ is the unit angular momentum vector of the gas averaged in a ring within the disc.

We will also examine the validity of the models in \ref{sec:validity} and verify that the wave-like regime is sufficiently resolved.

\subsection{Binary mass ratio}
\label{sec:massratio}
\begin{figure*}
    \centering
    \includegraphics[width=\textwidth]{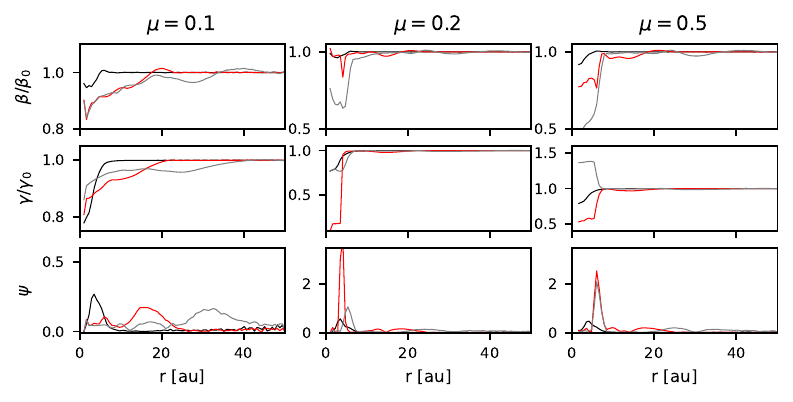}
    \caption{The tilt, twist and warp amplitude profiles for three values of the mass ratio $\mu$ for an inclination $i=45^\circ$ for $\mu=0.1$ and $0.2$ and $i=40^\circ$  for $\mu=0.5$. Black, red and grey lines denote 30, 200 and 500 binary orbits respectively. Note the different scales.}
    \label{fig:mass_ratio}
\end{figure*}

The mass ratio of the binary is expected to affect the evolution of a misaligned disc because it is a key factor contributing to the precession torques. Fig.~\ref{fig:mass_ratio} shows snapshots of the tilt and warp amplitude profiles as the disc evolves. For $\mu=0.1$ the disc develops a gentle warp with the initial misalignment of $45^\circ$, at which the precession torque is maximum. The warp amplitude profiles show the warp wave propagating outwards through the disc. As expected, we see {\edit stronger warping effects} with a larger mass ratio (as found by \citealt{facchini2013}), {\edit with the discs around the $\mu=0.2$ and equal mass binaries exhibiting disc tearing}.\footnote{{\edit We note that an exact definition of when a disc breaks is somewhat subjective. A rigorous definition would correspond to the stability analysis of \cite{dogan2018}, i.e. that the disc warp amplitude undergoes exponential growth (as shown to occur in numerical simulations by \citealt{raj2021a}). However, it is useful to have a definition that doesn't depend on the time evolution of the disc warp. Here we deem a disc to have broken if there is a sharp (i.e. radial scale of order $H$) variation in the disc tilt or twist that results in a large warp amplitude, accompanied by a strong depression of the local surface density of the disc \citep{nixon2012apr}.}} {\edit These results are consistent with \cite{ballabio2021}.}

The centre of the disc becomes twisted as well as tilted. When the disc breaks, the orientation of the outer disc remains unchanged from the initial conditions while the inner disc precesses independently. For the broken cases, the tilt and twist are largely constant outside the break compared to the case where the disc remains coherently warped (see Fig.~\ref{fig:mass_ratio}). {\edit For the $\mu=0.1$} model, the disc is beginning to evolve to a constant orientation. The inner region of this disc initially twists as well as tilts, launching the warp wave outwards, before evolving towards a flat, untwisted structure. 

It is clear that the disc with the {\edit $\mu=0.1$ binary (inclined by $45^\circ$) }remains stable: while the warp amplitude rises initially and shows a peak, over time this peak begins to decay and there is no clear sign of a break in the tilt or twist profiles. For the unstable discs with {\edit $\mu=0.2$ and $0.5$}, the warp amplitude increases rapidly and the disc breaks. For the stable disc, the location of the peak warp amplitude $\psi_{\rm{max}}$ moves steadily outwards. For the broken discs, $\psi_{\rm{max}}$ is located at the break and remains {\edit close to its initial} location.

\subsection{Binary-disc misalignment}
\label{sec:misalignment}
\begin{figure*}
    \centering
    \includegraphics[trim= 0cm 6cm 9cm 0cm, clip,width=\textwidth]{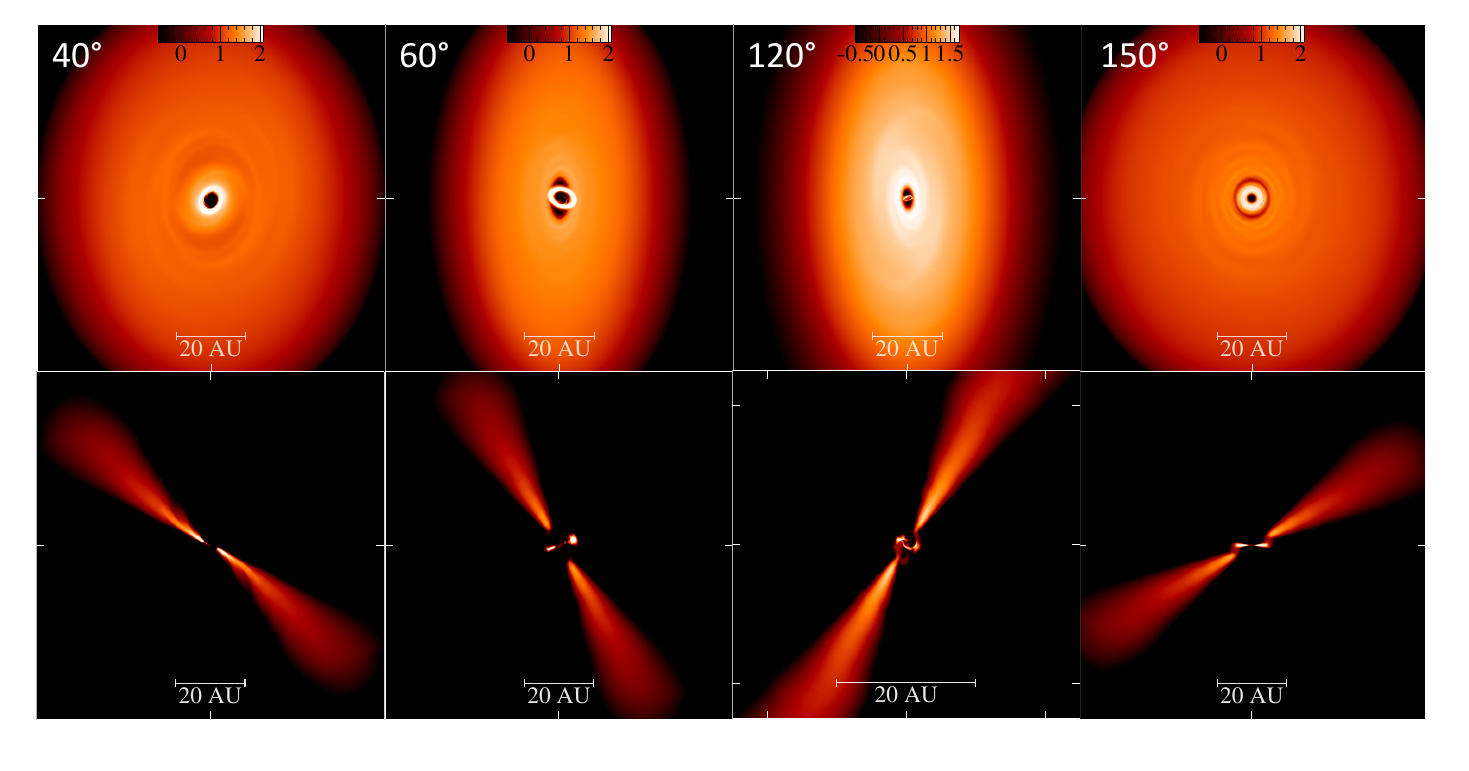}
    \caption{The range of structures produced in discs with different initial misalignments to the binary plane. The simulations all have $\mu=0.3$ and the initial misalignment is given at the top of each panel. Top row: Column density renderings along the $z$ direction. Units are $\log (\rm{g}~ \rm{cm}^{-2})$. Bottom row: density slices through the disc. Misaligned inner discs form more readily in retrograde discs and these tend to be smaller than in the prograde cases. Once the disc is broken, less warping of the coherent portions of the disc is observed when compared to the discs that remain unbroken.}
    \label{fig:misalignment_renders}
\end{figure*}

The effect of the disc inclination on the warp amplitude has been studied extensively for black hole accretion discs. For small inclinations, simulations demonstrated that the disc becomes warped, while for higher inclinations, ($i\gtrsim 30\degree$), rings tear off the inner edge of the disc and precess independently \citep[e.g.][]{nixon2012}. We now examine the structures that form for a protoplanetary disc that is misaligned with respect to the central binary and the dependence on misalignment angle.

In Fig.~\ref{fig:misalignment_renders} we present snapshots from simulations with different initial misalignments, all with $\mu=0.3$ and the same initial disc structure. With an initial misalignment of $i=40^\circ$ the disc develops a warp. In the left hand panels of Fig.~\ref{fig:misalignment_renders} we can see there is a region of decreased surface density in a ring at $R\approx10$~au, corresponding to the location of the maximum warp amplitude. Here the warp is sufficiently strong to affect the local surface density profile, but not strong enough to cause the disc to break. {\edit This depression of the surface density in a strongly warped, but unbroken, disc is a well-known feature of warped discs \citep[see e.g.][]{pringle1992,lodato2006,nixon2012apr,tremaine2014,nealon2022}.}

For $i=60$, 120 and $150^\circ$, the discs break and the outer disc warp is less prominent; the inner broken part of the discs have formed more closely aligned to the binary plane than to the outer disc. Comparing the $i=60^\circ$ and $i=120\degree$ discs, which are both equally offset from the binary orbital plane, the retrograde disc forms a smaller, less massive inner disc and has more prominent spirals. {\edit The inner disc for $i=120\degree$ has the form of an unstable ring with non-axisymmetric surface density, that precesses rapidly after tearing off from the outer disc.} Whereas the $i=40\degree$ disc maintained a smooth warp, the $i=150\degree$ disc breaks, illustrating a difference in behaviour between prograde and retrograde discs (we discuss this further in Section~\ref{sec:inneredge}). Additionally, we see accretion streams joining the inner and outer discs in the retrograde cases but for $i=60\degree$ the inner disc appears more disconnected; the strength of these structures depends on the phase of the precession of the inner disc.

\begin{figure}
    \centering
    \includegraphics[width=\columnwidth]{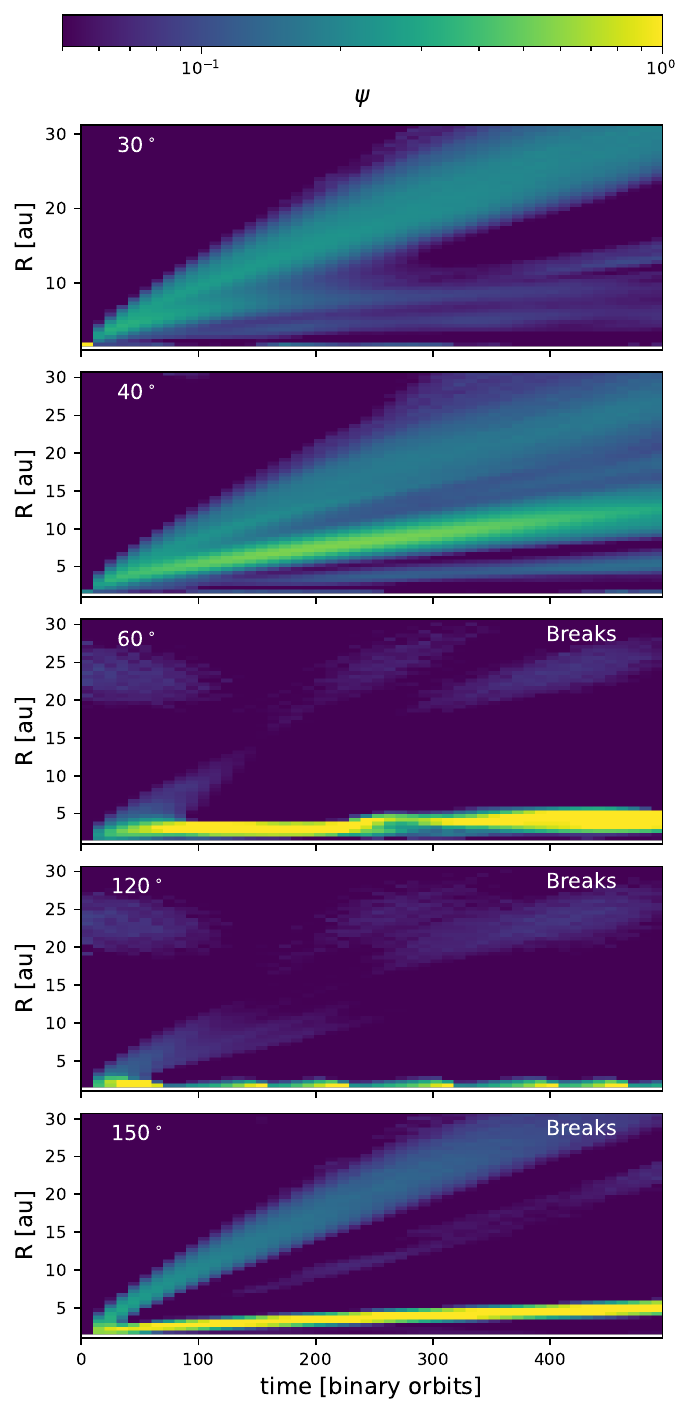}
    \caption{Evolution of the warp amplitude for discs with different initial misalignments and all with $\mu=0.3$. The three discs that break are labelled in the top right of the panels, and the location of the break can be seen as a strong (yellow) horizontal band marking a region of high warp amplitude. For an initial misalignment of $i=30^\circ$ the warp wave propagates freely through the disc and for $i=40^\circ$ the warp amplitude is starting to grow close to the inner edge. For $i=60^\circ$ and $120^\circ$ the warp wave barely propagates through the disc and the warp grows close to the inner edge, which causes the disc to tear. For $150^\circ$, a small warp propagates through the disc and the amplitude also increases near the inner edge where the disc exhibits a break.}
    \label{fig:misalign_psi}
\end{figure}

The resulting structures are driven by the propagation of the warp through the disc. Fig.~\ref{fig:misalign_psi} shows the time evolution of the warp amplitude $\psi$ for discs with various values of initial misalignment, zoomed in to show $R<25$~au. The wave warp is launched at the inner edge and propagates outwards. For $i=30$ and $40^\circ$, this figure shows strong warp waves propagating outwards, while for larger misalignments the warp amplitude rises close to the inner edge and there is little propagation through the disc. This is where the disc tears: $\psi$ grows rapidly and the disc becomes unstable. For the retrograde $i=150^\circ$ disc a weak warp wave propagates outwards through the disc while the inner region contains a break. In the discs that break, $\psi$ peaks close to the inner edge at a roughly fixed radius, and this marks the location of the break (see also Fig.~\ref{fig:misalignment_renders}). The location of the break is not exactly constant with time; for example in the $150^\circ$ case the break slowly moves outwards. The $i=120\degree$ disc shows somewhat periodic behaviour in the location of the peak in $\psi$. This is caused by accretion of the innermost ring: a ring gradually tears away from the inner edge, precesses and then collapses into the central cavity leaving a slightly increased inner edge to the circumbinary disc. As more material moves inwards, the ring forms again and the process repeats.

The precession torque driving the warping is $G_p \propto \cos(i)\sin(i)$ \citep[e.g.][]{nixon2013} and therefore the precession torque is maximised for a disc-binary inclination of $i=45^\circ$ and $i=135^\circ$. The results of the simulations are consistent with this, with misalignments close to $45^\circ$ and $135^\circ$ being more likely to exhibit disc breaking. The precession torque at any radius is the same for prograde and retrograde discs. However the evolution of the discs is different. Retrograde discs are more susceptible to breaking, and their warps and inner discs are more compact. We attribute this to the difference in the location of the inner edge of the circumbinary disc as a function of inclination angle. Retrograde discs are subjected to weaker resonances from the binary \citep[cf.][]{nixon2011,lubow2015,nixonlubow2015} and therefore the location of the inner disc edge is smaller for retrograde discs. A smaller inner edge to the disc allows the binary to exert a larger total torque, resulting in stronger warping of these discs.

\subsection{Sound speed profile}

The sound speed relates to the temperature profile of the disc via $c_s \propto \sqrt{T}$. The sound speed in the disc is set by a locally-isothermal equation of state with $c_s \propto R^{-q}$ and changing the value of $q$ determines how steeply the sound speed (and thus temperature) decreases with radius. {\edit In the limit of a flat disc (constant $H/R$),~ $q=0.5$. Protoplanetary discs are flared due to the thermal balance of stellar heating and local cooling, and calculations estimate such discs to have $q =0.2$ to 0.3 \citep{chiang1997,dalessio1998}. We compare the evolution of discs with $q=0.25$ (flared) and $0.5$ (flat surface).}

\begin{figure}
    \centering
    \includegraphics[width=\columnwidth,trim=0cm 0cm 0.5cm 0cm,clip]{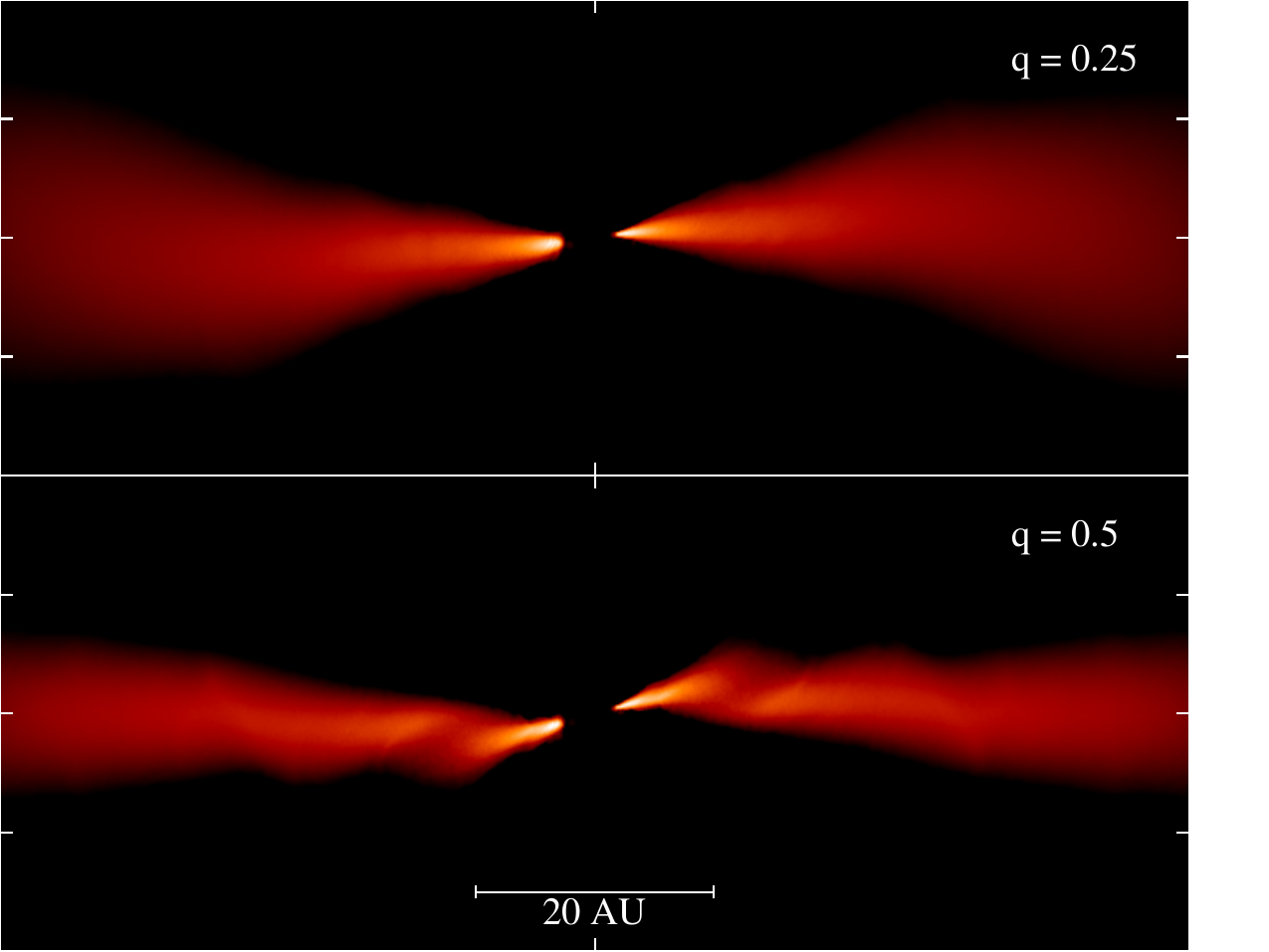}
    \caption{Density slices for different sound speed profile exponents, $q$, after evolving for 500 binary orbits. The initial inclination was $i=30^\circ$ and $\mu=0.3$. The difference in structure is clear: for $q=0.25$ the disc density profile is smooth and only slightly warped, but for $q=0.5$ the inner disc ($r\lesssim 15$~au) becomes misaligned to the outer disc and the density profile varies noticeably with radius.}
    \label{fig:c_s_slice}
\end{figure}

Fig.~\ref{fig:c_s_slice} shows the structure of discs with the two values of $q$. The $q=0.25$ disc is smoothly warped and flared and is noticeably thicker than the $q=0.5$ disc, since the scale height $H(R) = c_{\rm s}(R)/\Omega(R)$. The $q=0.5$ disc develops a misalignment in the inner $\sim 15$~au. This region never breaks off to form a separate inner disc but this warp is still likely to cast a shadow across the outer disc. These results agree with \citet{foucart2014} who found greater warping in a disc with $q=0.5$ than $q=0$.

\begin{figure}
    \centering
    \includegraphics[width=\columnwidth]{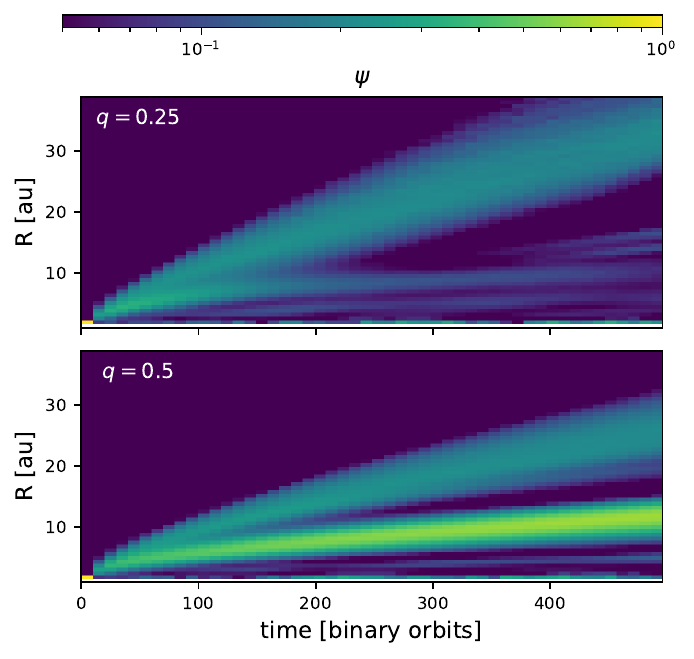}
    \caption{Evolution of the warp amplitude $\psi$ for discs with sound speed profile exponents $q=0.25$ and $0.5$. There is a clear difference in how quickly the warp propagated through the disc and also in the maximum value of $\psi$.}
    \label{fig:c_s_psi}
\end{figure}

We can understand the behaviour of the two discs by looking at how the bending waves can propagate through them. Bending waves propagate at a speed of $\sim c_{\rm s}/2$. The effect of this can be seen in Fig.~\ref{fig:c_s_psi}. The warp propagates more quickly through the $q=0.25$ disc when the sound speed profile is shallower and the warp propagation speed decreases less steeply with radius. In the $q=0.5$ disc, the maximum warp amplitude increases and this corresponds to the tilted inner region that develops.

The disc temperature, or sound speed, profile has a significant effect on the structure of a disc subject to a misaligned central binary or similar perturbation. These results imply that a disc that is susceptible to breaking may be stabilised by a shallower temperature profile and, conversely, a sharp change in temperature {\edit perhaps due to shadowing \citep[e.g.][]{casassus2019,keyte2023}}, could lead to a growth in warp amplitude.

\subsection{Surface density profile}
\label{sec:surfdens}

\begin{figure}
    \centering
    \includegraphics[width=\columnwidth,trim=0cm 3cm 0cm 0cm,clip]{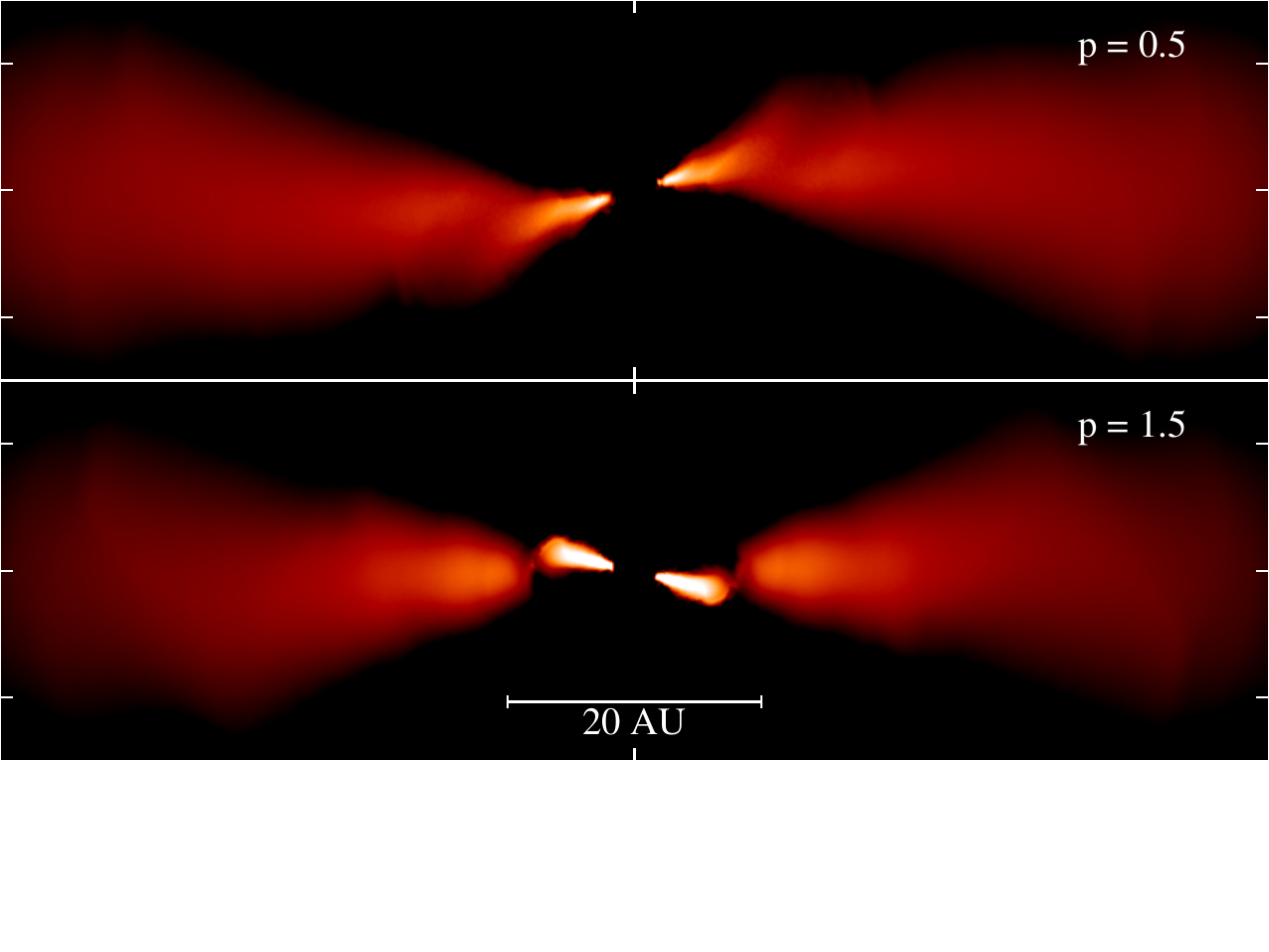}
    \caption{Density slice for discs with different initial surface density profile, $\Sigma \propto R^{-p}$ after $400$ orbits. The disc with the steeper density profile forms a separate inner disc while the disc with the shallower density profile remains stable.}
    \label{fig:Sigma_slice}
\end{figure}

The disc surface density is set via $\Sigma = \Sigma_0 R^{-p}$ and next we examine the effect of changing the surface density profile exponent $p$ on the disc stability. We performed two simulations of similar discs ($\mu=0.3$, $i=40^\circ$, $m_{\rm {disc}}=0.01$~M$_\odot$) with $p=0.5$ and 1.5. Density cuts for these simulations are presented in Fig.~\ref{fig:Sigma_slice}. While the $p=0.5$ disc develops a central warp and remains stable, a separate inner disc forms in the $p=1.5$ simulation where the disc breaks.

There are multiple effects to consider here. Firstly, the precession torque at a given radius is proportional to the surface density (equation~\ref{eq:Gp}). The total disc mass is the same in both cases so the disc with a steeper radial density profile has a greater mass in the inner region where the precession torque is stronger. For $p=0.5$, $G_p \propto R^{-3.5}$ and for $p=1.5$, $G_p \propto R^{-4.5}$. The torque driving the warp is therefore stronger in the disc with the steeper density profile \citep[see also][]{foucart2014}. When considering viscous discs, the increased precession torque is cancelled by the increase in viscous torque \citep[see e.g.][]{nixon2013}, however here the relevant effect is the communication of the warp via bending waves. Assuming the warp propagates with little dissipation, the amplitude of the propagating warp is expected to vary as it travels radially as $\psi(R) \propto \Sigma^{-1} R^{-3/2}$ \citep{nixon2010}. Consequently, with $p=0.5$, and thus $\psi(R) \propto 1/R$, the amplitude of a warp propagating outwards through the disc decays with time. However, for $p=1.5$ the warp amplitude is roughly conserved. This allows the warp amplitude to remain high for longer in the $p=1.5$ case, increasing the likelihood that the disc can achieve the necessary warp amplitude to break. Similarly we would expect a disc with $p>1.5$ to be more unstable as in this case the warp amplitude of a travelling wave can grow as it propagates outwards away from the binary.\footnote{Note that for misaligned discs around the primary or secondary star in a binary, such as those simulated by \cite{dogan2015}, the torque from the binary is applied primarily to the outer disc edge, and in this case smaller values of $p$ are more likely to result in instability of the disc warp.} {\rev We defer a detailed discussion of this effect to a future paper.}

\begin{figure}
    \centering
    \includegraphics[width=0.9\columnwidth]{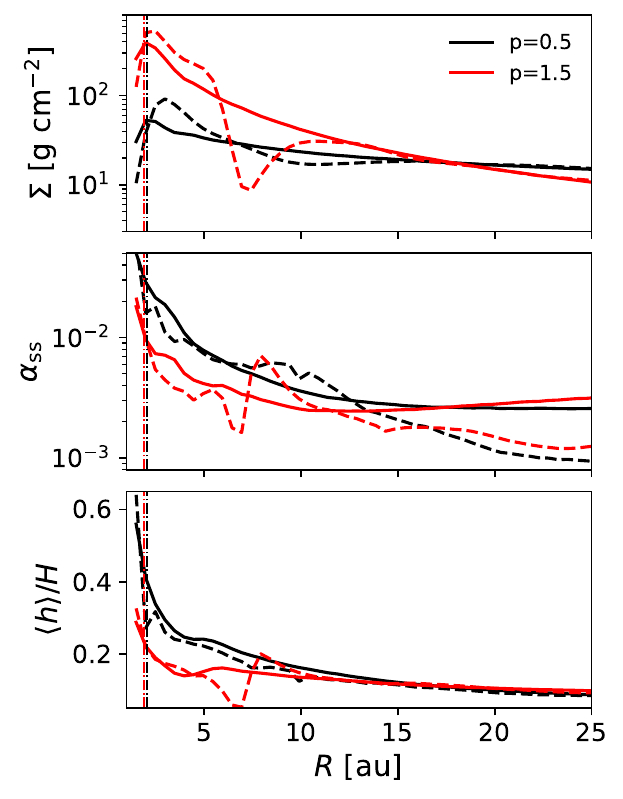}
    \caption{The surface density profile $\Sigma$, estimated effective viscosity $\alpha_{\rm{ss}}$ and vertical resolution for the simulations with surface density profile exponents $p=0.5$ and 1.5 after 30 binary orbits (solid lines) and 400 binary orbits (dashed lines). The vertical dash-dotted lines indicate the locations of the inner edge for the two density profiles after 400 binary orbits (See section \ref{sec:inneredge} for definition). }
    \label{fig:alpha_Sigma}
\end{figure}

Secondly, due to the nature of the SPH method, the change in mass distribution introduces a difference in resolution and therefore also in the magnitude of the numerical viscosity arising from the SPH artificial viscosity terms. {\edit To assess the possible impact of the difference in resolution we estimate the effective $\alpha_{\rm ss}$ following \citet{meru2012}:
\begin{equation}
\label{eq:alpha_ss}
     \alpha_{\rm ss} = \frac{31}{525}\alpha_{\rm SPH}\left(\frac{\langle h \rangle}{H}\right) + \frac{9}{70\pi}\beta_{\rm SPH}\left(\frac{\langle h \rangle}{H}\right)^2.
\end{equation}
As explained in Section \ref{sec:hydro}, $\alpha_{\rm SPH}$ is varied between 0 and 1 by the switch of \citet{cullen2010}. We take the average of $\alpha_{\rm SPH}$ in concentric spherical shells to estimate $\alpha_{\rm ss}$. $\langle h \rangle$ is the shell-averaged smoothing length and $H$ is the disc scale height. The vertical density profile of a protoplanetary disc follows a Gaussian distribution. The scale height can therefore be estimated from the SPH simulations after binning particles into shells:
\begin{equation}
\label{eq:SPHscaleheight}
    H(R) = \sqrt{\frac{1}{N(R)} \sum_{i=1}^{N(R)} (z'_i-\bar{z'})^2}.
\end{equation}
$N(R)$ is the number of particles in the bin corresponding to radius $R$ and $z_i' = \hat{l}\cdot \vec{x}_i$ is the height of the particle at position $\vec{x}_i$ above the (warped) disc midplane (and $\hat{l}$ is the unit angular momentum vector of the disc at radius $R$).

} The radial profiles of the surface density, and the resulting effective $\alpha_{\rm{ss}}$ and vertical resolution for both discs are presented in Fig.~\ref{fig:alpha_Sigma}. Since $G_{\rm p }\propto \Sigma$, the precession torque is nearly an order of magnitude greater for the $p=1.5$ disc and we expect this stronger torque to drive a larger warp. Additionally, the $p=1.5$ disc has a smaller cavity (see section \ref{sec:inneredge} for further discussion) which would also drive a larger warp due to having material closer to the stars, since $G_{\rm p }\propto R^{-(p+3)}$. We need to verify whether the difference between the models is due to physical or numerical effects. The inner edge of the $p=1.5$ disc is better resolved (see Fig.~\ref{fig:alpha_Sigma}) and as such has a lower value of $\alpha_{\rm{ss}}$ at the inner edge. \citet{facchini2013} showed that poorer resolution leads to a larger truncation radius and therefore a smaller torque on the disc. Here, the $p=1.5$ disc has a slightly smaller cavity (after 400 orbits the difference is $R_{\rm in}=2.1$ vs $1.9$~au for $p=0.5$ and $p=1.5$ respectively). This means that the effect of changing the surface density profile by varying $p$ produces a significantly larger (physical) effect than the (numerical) effect produced by a change in inner disc location. The former effect is at the order of magnitude level, whereas the latter is at the level of a factor of $(2.1/1.9)^3 \approx 4/3$.

\subsection{Disc outer radius}
\label{sec:r_out}
We now examine the effect on the warp evolution of a reduced disc outer radius, $R_{\rm out}$, on a disc with $i=30\degree$ and $M_{\rm disc} = 0.01$~M$_{\odot}$. The baseline model had $R_{\rm out}=50$~au and the models with a reduced outer radius have $R_{\rm out}=40$ and 30 au respectively. The initial conditions were created by removing particles outside their respective $R_{\rm out}$ from the initial file of the $R_{\rm out}=50$~au {\edit simulation} to ensure the density profile was otherwise identical.

\begin{figure*}
    \centering
    \includegraphics[width=0.9\textwidth]{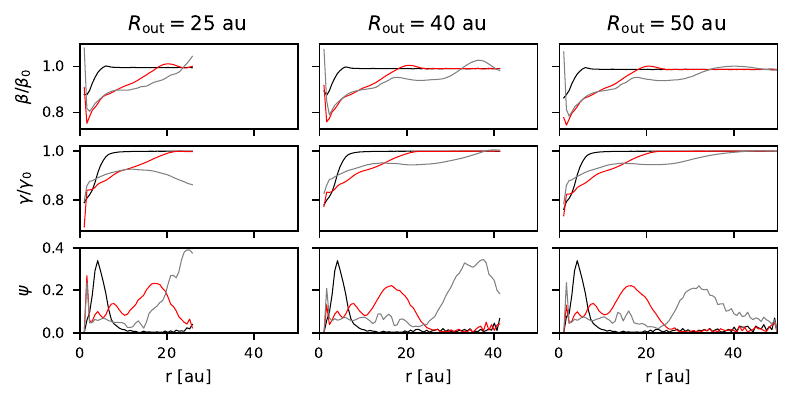}
    \caption{The tilt, twist and warp amplitude profiles for discs of different outer radius. Black, red and grey lines denote 30, 200 and 500 orbits respectively. By 500 orbits, the warp wave has reached the outer edge of the smallest disc and reflected back inwards, causing differences in the disc structure.}
    \label{fig:rout_betapsi}
\end{figure*}

While the initial misalignment between the disc and binary planes was identical in each case, the reduction of the disc radius decreases the angular momentum of the disc and so the angle between the disc angular momentum vector $\hat{L}_{\rm{disc}}$ and the net system angular momentum vector $\hat{L}_{\rm{tot}}$ is increased. The disc + binary systems evolve towards alignment, such that $\hat{L}_{\rm{disc}}\parallel \hat{L}_{\rm{binary}} \parallel \hat{L}_{\rm{tot}}$. With a smaller fraction of the total angular momentum held in the disc, the smaller disc must change its tilt the most to become aligned with its host binary.

\begin{figure*}
    \centering
    \includegraphics[width=0.9\textwidth, trim=0cm 6.5cm 0cm 0cm,clip]{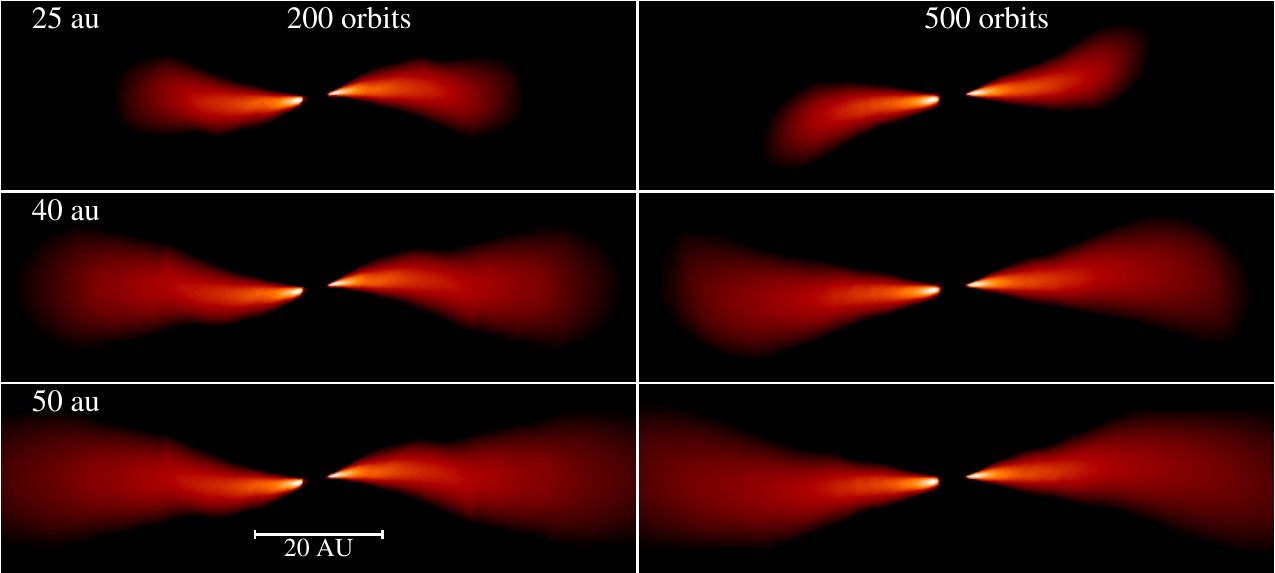}
    \caption{Density slices to show the morphology of discs with different outer radii after 200 and 500 binary orbits.}
    \label{fig:rout_slices}
\end{figure*}

The communication timescales, the time for bending waves propagating at $c_{\rm s}/2$ to traverse the disc, are 500, 370 and 190 years (710, 520, and 270 binary orbits) for $R_{\rm out}= 50$, 40 and 25~au respectively. These values were found by integrating the sound speed profile from $R_{\rm in}$ to each value of $R_{\rm out}$:

\begin{equation}
\label{eq:tcomm}
t_{\rm{comm}}\approx \int _{R_{\rm in}} ^{R_{\rm out}} \frac{2}{c_{\rm s}(R)} dR.
\end{equation}

The evolution of $\psi_{\rm max}$, shown in Fig.~\ref{fig:rout_betapsi}, is similar for all three discs until the warp wave reaches the edge of the smallest disc. Around this time the outer regions of the disc are affected by reflections of the bending waves from the edge. This may be less pronounced in discs where the density drop at the outer edge is more gradual. The tilt evolution is very similar for $R\lesssim20$~au and by 500 binary orbits none of the discs is close to alignment. The smallest disc has been twisted the most and shifted from its initial phase.

Before the initial warp wave reaches the edge of the 25~au disc, the density slices confirm that the morphology of the disc is similar for all of the model discs (see Fig.~\ref{fig:rout_slices}, left column). However, later, after 500 binary orbits, the smallest disc becomes warped throughout, with both tilt and twist varying with radius, since it lacks the anchoring effect of an outer disc. The large warp seen early on is transitory and the longer term evolution of the disc morphology will depend upon the reflected bending waves. The $R_{\rm{out}}=25$ au model displays warping at the edge of the disc where this is beginning to develop. 

\subsection{Binary eccentricity}
\begin{figure*}
    \centering
    \includegraphics[width=\textwidth]{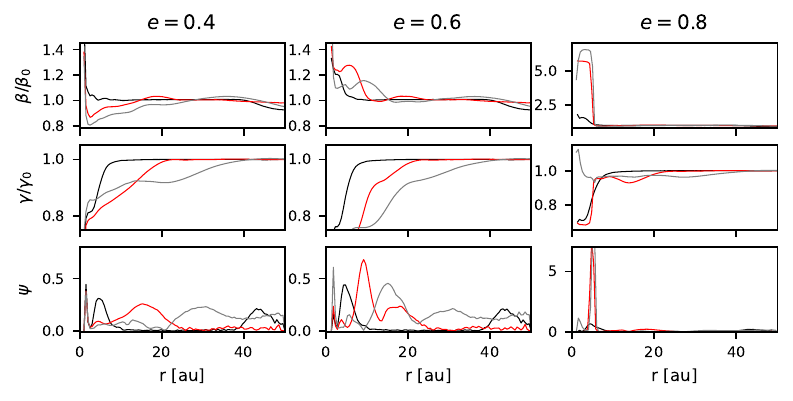}
    \caption{The effect of eccentricity: a disc initially inclined at $25^\circ$ to the binary, with different values of binary eccentricity, $e$. Black, red and grey lines denote 30, 200 and 500 binary orbits respectively. Note the different scales for the right hand column.}
    \label{fig:ecc}
\end{figure*}

When the binary eccentricity is small, circumbinary discs tend to precess around the binary angular momentum vector as described above. However, if the binary eccentricity and the disc-binary inclination is sufficiently large then the precession can be predominantly around the binary eccentricity vector instead. For a fluid disc, this results in alignment of the disc to the plane defined by the binary eccentricity vector rather than the binary orbital angular momentum vector. This process was derived analytically and confirmed numerically by \cite{aly2015}, and was subsequently applied to the dynamics of protoplanetary discs by \cite{martin2017}. \cite{aly2015} also showed that the precession induced by an eccentric binary can lead to disc tearing. Their simulations employed thin discs and were focused on the diffusive regime of warp propagation.

\citet{stevenson2022} presented some simulations of small protoplanetary discs around an eccentric binary, showing that a more eccentric binary caused the disc to warp more and that a highly eccentric binary ($e=0.8$) caused the disc to tear. Here we examine further whether an eccentric binary can tear a low inclination wave-like disc that is stable for a circular binary and we explore how the system evolves over several hundred binary orbits.

We performed simulations of a disc with $i=25^\circ$ and $\mu=0.5$ for $e=$~0.4, 0.6 and 0.8. The lower values of eccentricity caused the disc to warp and for $e=0.8$ the binary was able to tear the disc, even with a relatively small misalignment, as seen in \citet{stevenson2022}. The inner regions of the disc are far less stable than those formed in simulations with larger misalignments. Accretion streams form and the inner disc evolves to a near-polar ring-like structure. After precessing as a coherent ring and reaching a roughly polar configuration, the ring becomes eccentric. Material from the ring is thrown against the inner edge of the outer disc. Gas accretes rapidly from the outer disc, feeding the inner ring, therefore we expect the inner disc to persist even though it is unstable because it is constantly replenished.

The tilt, twist and warp amplitude profiles for the discs around eccentric binaries are presented in Fig.~\ref{fig:ecc}. For an initial eccentricity of $e\geq0.6$, disc tilt is increased in the opposite direction (compare also with Fig.~\ref{fig:mass_ratio}). \citet{aly2015} described this effect: for the lower eccentricity the inner disc is aligning with the binary but for higher values of eccentricity, the disc precesses about the eccentricity vector and evolves to polar alignment due to dissipation. The chaotic evolution of the inner ring is not dissimilar to that found by \citet{aly2015}, despite the difference in the disc viscosity (they use $\alpha_{\rm{ss}}=0.1$ and $H/R = 0.01$ at the inner edge, placing the disc firmly in the viscous regime). With the high eccentricity of the inner regions of the disc, the evolution is highly dissipative, regardless of the viscosity and disc thickness so we may expect the evolution to be similar.

 \subsection{Significance of the location of the inner disc edge}
\label{sec:inneredge}

\begin{figure}
    \centering
    \includegraphics[width=\columnwidth]{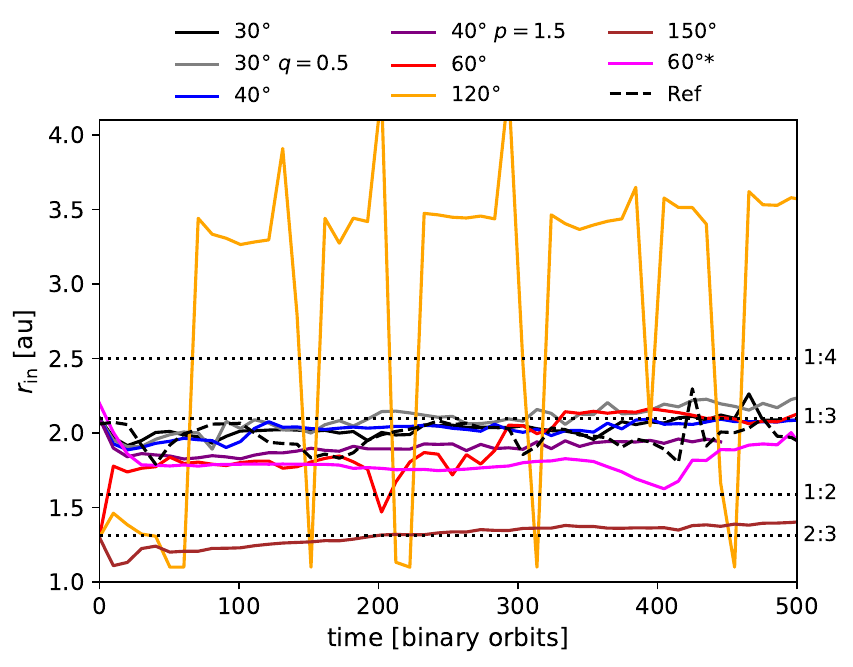}
    \caption{Locations of the disc inner edge for selected simulations with $\mu = 0.3$, including the reference model with the binary and disc aligned. Locations of outer Lindblad resonances are indicated. In the $i=120\degree$ disc, an inner ring forms periodically and collapses onto the stars, clearing an inner cavity. The simulation labelled ``60$\degree$*'' has initial inner radius $r_{\rm{in}}=2.2$~au. In this simulation the disc also breaks, but forms a slightly radially-thicker inner disc than the $60\degree$ simulation.}
    \label{fig:Rin_LR}
\end{figure}

The location of the inner edge of the disc is typically set by tidal truncation at the innermost outer Lindblad resonance where the tidal torques exceed the viscous torques that drive the disc to spread inwards \cite[][see also the discussion in \citealt{heath2020}]{artymowics1994}. The precession torques that drive warping are a strong function of radius, decaying rapidly with distance from the binary. The presence of gas at smaller radii as (dis)allowed by tidal truncation therefore affects the possible strength of the torques driving the warping. For this reason, we present the locations of the disc inner edge for selected simulations in Fig.~\ref{fig:Rin_LR}. The inner edge is defined as the location at which the shell-averaged density falls to half of the maximum value. The density profile at the inner edge is highly variable, especially for discs which have formed an eccentric inner ring or spiral accretion streams.

The initial locations of the disc inner edge at the start of the simulations were set as described in Table \ref{tab:sim_parameters}. These were chosen to be close to the expected inner edge location for each binary mass ratio and inclination in question from the analysis of \citet{miranda2015} so that the disc inner edge would settle into a steady state reasonably quickly. To investigate the effect of the initial value on the subsequent location of the inner edge, the $\mu=0.3$ $i=60\degree$ simulation was repeated but with $R_{\rm {in}} = 2.2$~au and this is included in Fig.~\ref{fig:Rin_LR}.

The inner edge moves inward initially as the initial conditions relax and the simulations with $i\leq 60\degree$ settle close to the 1:3 commensurability (corresponding to the $(m,l)=(2,1)$ outer Lindblad resonance). The central cavity of the disc with the steeper surface density profile ($p=1.5$) is slightly smaller, and steepening the sound speed profile ($q=0.5$) widened the cavity. The disc with initial $i=120\degree$ periodically forms an inner ring which then collapses in, briefly leaving a wider cavity before material accretes inward again.

The evolution of the $i=60\degree$ simulations appears to depend to some extent on the initial location of $r_{\rm in}$. After $\sim 50$ orbits, the inner radii of both simulations are similar but then they deviate again after $\sim 200$ binary orbits. In terms of the disc morphology, both discs break but the disc with the larger initial $R_{\rm in}$ tears off a larger inner disc roughly $100$ binary orbits later (see Section \ref{sec:rbreak}). Although the inner radii tend to remain within $0.5$~au of their initial value the subsequent evolution is somewhat sensitive to this initial value; {\edit this demonstrates that the torque applied to the disc is strongly dependent on the distribution of mass near the disc inner edge.}

From Fig.~\ref{fig:Rin_LR}, the inner edge of the $i=60\degree$ disc does seem to lie at the 1:3 resonance but takes a long time to settle. It is worth noting that the quoted inclination values for each disc are the initial values, and as the simulation progresses the inner disc inclination typically decreases as the disc aligns to the binary orbital plane.

The locations of the outer Lindblad resonances provide a fair starting point for the simulations but we do see a shift in the inner edge location. The calculations of \citet{miranda2015} involve several assumptions that result in discrete values for truncation radii as a function of inclination and eccentricity. In real (and simulated) discs, Lindblad resonances have a finite width and this is likely to account for the variations we observe.


\subsection{Where does the disc break?}
\label{sec:rbreak}

\begin{figure}
    \centering
    \includegraphics[width=\columnwidth]{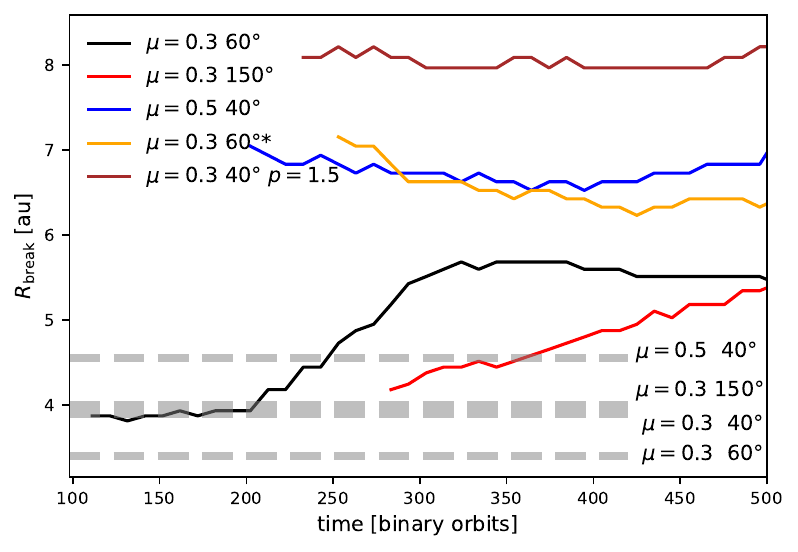}
    \caption{The radius of the break in the disc as a function of time, commencing when the disc first broke, in spherical coordinates. Dashed grey lines indicate the critical radius predicted from the intersection of the precession timescale and  warp communication timescale for each simulation {\edit using the method described in Section \ref{sec:rbreak}}. From those simple theoretical arguments, we would expect the disc to be stable for $R>R_{\rm{crit}}$ but the simulated discs break further out than predicted. The simulation marked with an asterisk is identical to $\mu=0.3$ $i=60\degree$ but with the inner edge initially set at 2.2~au rather than 1.3~au.}
    \label{fig:rbreak}
\end{figure}

The question of where an unstable disc breaks is central to interpreting observations of broken discs. It is by now clear that many factors affect the stability of a protoplanetary disc to breaking so is it possible to predict the breaking radius?

While their main focus is diffusive black hole accretion discs, \citet{nixon2013} provide a rough estimate for where wave-like discs might be expected to break based on the simple argument that to build up a substantial warp the disc must be forced to precess faster than warp waves can propagate. This suggests that discs should break at radii smaller than a critical value $R_{\rm crit}$ given by their equation~A3. The discs simulated here are flared (i.e. $H/R$ increases with radius) and so we calculate the breaking criterion for a representative range of $H/R$ values. The values we obtain lie $1.4 < R_{\rm{crit}} < 2.7$~au. However, the four discs broke (as determined by the tilt and twist profiles, and the accompanying sharp drop in density) at $R\gtrsim 4$~au (Fig.~\ref{fig:rbreak}), up to $\sim 8$~au. (We exclude the $\mu=0.3$, $i=120\degree$ disc from this analysis because the inner region periodically collapses into the stars, which makes a breaking radius difficult to define.) This suggests that the equation derived by \citet{nixon2013} is at best a conservative estimate, and indeed \citet{nixon2013} caution that it does not account for several physical effects. Since the predicted breaking radii are very different from those found in the simulations we caution against using this simple criterion. 

It is expected that a break occurs where the disc is unable to communicate the warp induced by precession efficiently such that the warp amplitude grows rapidly \citep[e.g. Section~5 of][]{dogan2018}. This may be where the precession torque exceeds the viscous torque in the case of diffusive discs. For wave-like discs, this is where the time scale for the propagation of bending waves is greater than the precession time. This follows the reasoning presented in \cite{nixon2013}, but here we improve on this by calculating the warp communication timescale more accurately. We calculate the communication timescale using equation \ref{eq:tcomm} for each simulation using the inner radius obtained in Fig.~\ref{fig:Rin_LR} at the time of breaking. {\edit The precession timescale is given by

\begin{equation}
\Omega_{\rm p} = \frac{3}{4} \mu \frac{a^2}{R^2} \Omega \cos(i), \\
t_{\rm p} = \left | \frac{2\pi}{\Omega_{\rm p}} \right |.
\label{eq:tprec}
\end{equation}

$R_{\rm {crit }}$ for these criteria is therefore the radius at which $t_{\rm p} = t_{\rm {comm}}$ and the} region $R<R_{\rm{crit}}$ is expected to be unstable to tearing, i.e. below the dashed lines in Fig.~\ref{fig:rbreak}. Once again, however, these values are smaller than the breaking radii observed in the simulations.

The location of the break can vary with time because these discs are typically not in a steady state. Recent theoretical work indicates that breaking occurs where the warp amplitude peaks \citep{dogan2018,deng2022}. These results were derived for a steady warp where these quantities are well defined and are not easily applicable to these simulations where the disc is subject to forced precession and the location of $\psi_{\rm max}$, for example, is evolving. After a few hundred orbits the breaking radius appears to reach a somewhat steady value in most cases, {\edit and this timescale is longer than the wave communication timescale to the break radius}. We note that the $i=150\degree$ disc remains stable for nearly 300 binary orbits before tearing, highlighting the importance of running the simulation for long enough when seeking to determine whether a disc is stable. {\edit The discs which broke did so at $t<t_{\rm{comm}}$, which sets a limit on the necessary simulation time.} Another point of note is that the two simulations for $\mu=0.3$, $i=60\degree$ with different initial $R_{\rm {in}}$ tear at different radii. Comparing with Fig.~\ref{fig:Rin_LR}, after 300 orbits, $R_{\rm{in}}$ has evolved such that the inner edge of the disc that was initialised with $R_{\rm {in}}=2.2$~au lies at a smaller radius that the disc with  $R_{\rm {in}}=1.3$~au.

In summary, there is a large variation in the radius at which a disc breaks and this radius may evolve as the structure of the disc evolves. It is clear that a simple estimate based on the wave communication and precession timescales does not adequately capture the range of radii over which these discs can tear, and we discuss the possible reasons for this further in section \ref{sec:discussion_factors}. While an {\it a priori} criterion for disc tearing is highly desirable, it seems that for circumbinary discs there is sufficient dependency on a wide range of parameters and components of the initial conditions that such a criterion remains elusive.

\subsection{Model validity}
\label{sec:validity}
\begin{figure}
    \centering
    \includegraphics[width=\columnwidth]{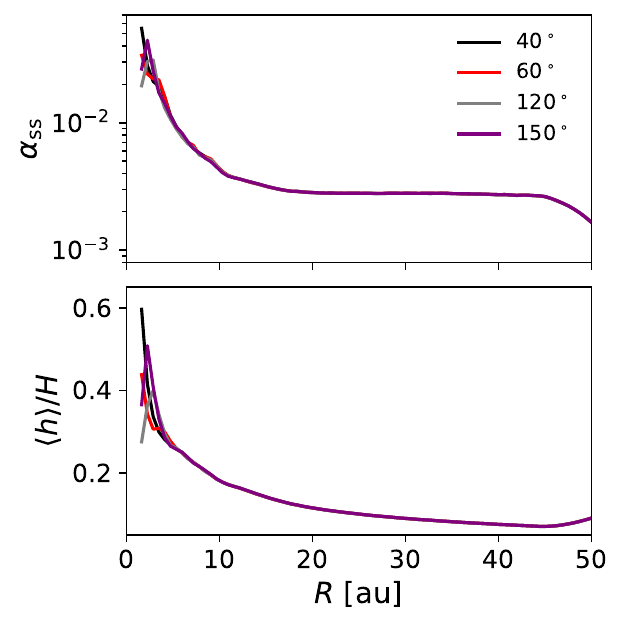}
    \caption{Top panel: The approximate value of $\alpha_{\rm {ss}}$, shown for the simulations of misaligned discs presented in \ref{sec:misalignment} after 30 binary orbits. Note that the value of $\alpha_{\rm{SPH}}$ is variable so the values here are averaged in concentric spherical shells. The lines are also cut off at small radii where $\rho$ falls to $\rho_{\rm{max}}/2$, which we take as delimiting the location of the disc inner edge. $H/R > 0.05$ for these simulations and so the wave-like condition is satisfied at all radii. Lower panel: the vertical resolution of the same simulations.}
    \label{fig:alpha_plot}
\end{figure}

{\edit The disc must be adequately resolved in the vertical direction to properly model breaking \citep[see][]{nealon2015,drewes2021} and we require} the effective viscosity $\alpha_{\rm ss} < H/R$ to adequately capture the dynamics of the disc in the wave-like regime. To assess whether the models are indeed in the wave-like regime we average $\alpha_{\rm SPH}$ in radial bins to estimate $\alpha_{\rm ss}$ using Equation~\ref{eq:alpha_ss}. In  Fig.~\ref{fig:alpha_plot}, we plot $\alpha_{\rm {ss}}$ for the four simulations from section \ref{sec:misalignment} after 30 binary orbits. In the lower panel we show the vertical resolution {\edit as a fraction of the local disc scale height, finding that $\left<h\right>/H \sim 0.1-0.3 \ll 1$ in the body of the disc, which is sufficient to resolve disc breaking.}

\begin{figure}
    \centering
    \includegraphics[width=\columnwidth]{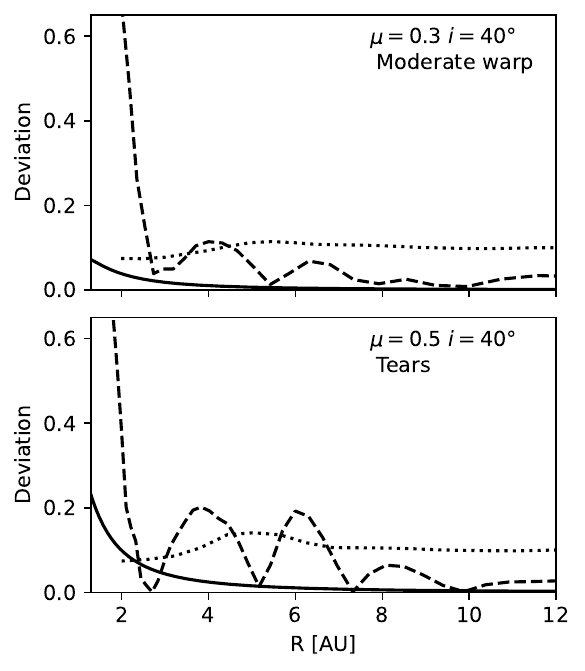}
    \caption{Verification that the simulations meet the near-Keplerian rotation criterion for wave-like warp propagation given by equation \ref{eq:kepcondition}. The left hand side of equation \ref{eq:kepcondition}, the deviation from Keplerian rotation, calculated from test orbits in the binary potential (solid lines) and the fluid orbits in the simulations (dashed lines). Dotted lines indicate $H/R$ taken from the simulations. The snapshots for both simulations in the figure were taken after 90 binary orbits, before the disc represented in the bottom panel tears. The disc in the top panel develops a moderate warp but remains stable. The Keplerian condition is satisfied throughout the disc for $\mu=0.3$ and in all but the innermost $\sim1$~au for the $\mu=0.5$ disc, where the deviation exceeds $H/R$. The sharp spike in the dashed line towards the inner edge of the disc is due to the strong density gradient in this region where the surface density falls rapidly towards the inner disc edge.}
    \label{fig:kep_condition}
\end{figure}

We see that the simulated discs satisfy the condition for wave-like evolution. Close to the inner edge, the artificial viscosity increases significantly and this region may stray into an intermediate regime. The main impact this could have is to shift the precise location of the inner edge of the disc. In Section \ref{sec:inneredge} we found that the inner edge did sit inside the 1:3 resonance for the first $\sim300$ orbits, which could potentially be due to the increased viscosity at the edge. Additionally, we have verified that the discs are nearly Keplerian, as required for the wave-like propagation of warps. Fig.~\ref{fig:kep_condition} shows the deviation from Keplerian rotation for two simulations, calculated from equation \ref{eq:kepcondition}. The solid lines indicate this condition calculated directly from the binary potential, following a similar method to \citet{dogan2020}. The criterion given by Eq.~\ref{eq:kepcondition} is satisfied ($<H/R$, dotted lines) for the binary potential for the whole disc for $\mu=0.3$ and in all but the innermost $\sim1$ au for $\mu=0.5$. The dashed lines indicate the deviation from Keplerian rotation calculated from the azimuthal gas velocity from the simulations. This shows that the pressure gradient in the disc due to the warp also results in a deviation from Keplerian rotation in addition to that due to the binary potential.

\section{Discussion}

\subsection{What factors affect the structure of the disc?}
\label{sec:discussion_factors}
The interaction of a circumbinary disc with its host stars is complex. As such, we find that no single parameter is responsible for the amplitude of the warp, which means we need to consider more than just the angle of misalignment for assessing the stability of a protoplanetary disc. We have shown, for example, that the temperature and density profiles can make the difference between a stable warp and a torn disc.

In the simulations, the disc evolution is sensitive to the initial location of the inner edge. Although we set $R_{\rm{in}}$ to be at the expected location of the outer Lindblad resonance responsible for clearing the cavity, the disc spreads inwards before settling near the original location after a few hundred {\edit binary} orbits. The two simulations set up with only the initial inner radius differing showed differences in their evolution and the size of the inner cavity did not converge after a few hundred binary orbits. For those two models, the location of $R_{\rm{in}}$ did not affect whether the disc tore, but did affect the breaking radius and how quickly the disc tore.

It is difficult to predict whether a disc will break because of the contributions from multiple factors and the sensitivity to the inner edge location. We can place some constraints on the conditions for which the disc will be stable from the simulations presented here. The simulations suggest that, for typical protoplanetary disc properties, a binary with a low mass ratio $\sim 0.1$ is unable to tear the disc. We find that the disc is also stable for a misalignment of $i\lesssim 40\degree$, and this increases for unequal mass binaries. However, a high binary eccentricity $e\gtrsim 0.8$ can destabilise a warped disc of low mass ratio or low misalignment. \citet{deng2022} found the disc broke at a much smaller angle of $14\degree$, albeit for a thinner disc, and state that the affine model predicts the disc will break for $i>20\degree$ for $H/R=0.05$. The discs in this paper are flared, such that $H/R > 0.05$, which increases the tilt angle required to break the disc. The effects of varying the sound speed profile or density profile only come into play for marginally stable discs, in which different values of these parameters can make the difference between a smooth warp and a broken disc. Given that misalignments are likely to be caused by irregular accretion from the surrounding molecular cloud, we can expect variations in surface density and the temperature profile, which could well affect how the warp propagates. 

The discs that break maintain a planar (unwarped) outer disc, for example the $\mu=0.5$ simulation in Fig.~\ref{fig:mass_ratio}, and this warrants further discussion. This is a very different structure to that of the stable {\edit $\mu=0.1$ disc} which is tilted and twisted right to the outer edge. From this we can infer that throughout most of the disc the warping is the result of the warp waves propagating outwards from the inner edge. If the precession torque were responsible for warping the outer regions of the disc, we would expect the outer disc of the {\edit $\mu=0.2$ and $0.5$ models} to be warped. However, the warp amplitude drops abruptly at the location of the break, indicating that communication of the warp is effectively halted, and that the precession torques are too weak at greater radii to drive a warp independently of the dynamical state of the inner disc. Fig.~\ref{fig:misalign_psi} shows this clearly: the panels for $i=30$ and $40^\circ$ show the warp propagating through the disc and we can see this is severely restricted for the more misaligned discs. For the latter, the wave propagation is inhibited and the warp amplitude grows, leading to the break.

The results of \citet{facchini2013} also show a flat outer disc after tearing has occurred and this is a clear difference in structure to that of torn black hole accretion discs. Black hole discs maintain smooth twist and tilt profiles after tearing and the outer disc is typically warped, except for the lowest values of black hole spin \citep{nixon2012,nealon2015}. This result has also been found for wave-like discs around a black hole \citep{drewes2021}. For a disc around a black hole the precession frequency is $\propto R^{-3}$, whereas for a circumbinary disc the precession frequency is $\propto R^{-7/2}$. This means that the precession torque decays more rapidly with increasing radius in the circumbinary case, compared to the black hole case. This difference may explain why the disc can develop a noticeable warp outside of the broken regions in the black hole case, and that the outer regions of circumbinary discs are comparably planar. In both cases, the torque is highly sensitive to the location of the inner disc edge, and the torque applied at $R\gg R_{\rm in}$ has only a small effect on the resulting disc structure.

When studying the breaking radius (section \ref{sec:rbreak}), we compared the precession and communication timescales in the disc in an attempt to predict where the disc would break, following \citet{nixon2013}. We can now understand why this method significantly underestimates the breaking radius. This method assumes that the structure of the disc at a given radius is driven by the precession torque at that location. However, the disc structure is dominated by the bending waves driven near the inner edge which propagate outwards. This suggests that a {\it local} comparison of the precession torque and the wave travel timescale (e.g. that presented in the appendix of \citealt{nixon2013}) is not appropriate for determining where the disc breaks.\footnote{{\edit It is worth noting that previous authors have found some level of agreement with the local wavelike criterion presented in the appendix of \cite{nixon2013}, see e.g. \cite{nealon2015,facchini2018}. However, \cite{drewes2021} have noted that these works present numerical simulations that are in the diffusive regime of warp propagation due to the magnitude of the numerical viscosity present in those simulations. The simulations we present here are in the wavelike regime (see Fig.~\ref{fig:alpha_plot} which accounts for the numerical viscosity), and thus it is not surprising that the results we obtain for where the disc breaks are distinct from, and comparatively larger than, previous works.}} {\edit Instead, it is likely that a global analysis is required to determine the stability of wavelike warped discs.}

\subsection{Limitations of the models}
\label{sec:limitations}

Like \citet{facchini2013}, our simulations produce a twist in the inner region of the disc. According to their 1-D analytical model \citep[see][]{lubow2002}, an inviscid disc should have a constant twist angle. When a viscosity is introduced, the disc can become twisted, with the exact shape depending on the magnitude of the viscosity. In the simulations presented here, the viscosity is purely numerical since we aimed to minimise the net viscosity as much as possible. The viscosity of protoplanetary discs is likely to be an order of magnitude or more lower than our simulations ($\alpha_{\rm {ss}}\lesssim 10^{-3}$, e.g. \citealt[][]{rosotti2023}) but to achieve a lower numerical viscosity would require an increase in resolution to at least $10^8$ SPH particles, which is currently computationally infeasible. From an observational perspective, tilt and twist amount to a single distortion of the disc. The so-called twist that is observed in velocity maps is a result of the continuous change in the slope of the midplane, radially and azimuthally, i.e. the warp. It was shown in \citet{young2022} that this is not a direct measure of the warp amplitude because it depends on the optical depth of the observed molecular line. While the viscosity of the simulations may increase the twist of the disc, this is probably less significant than the additional twist observed due to optical depth effects.

The simulations presented here are isothermal, like the majority of prior work. However, there is likely to be significant local heating at locations where the warp amplitude grows sharply, which could stabilise the disc. \citet{deng2022} have shown that in adiabatic models the inner and outer parts of a broken disc reconnect smoothly. A broken inner disc could be a short-lived state and, as the disc components reconnect, the disc will evolve towards alignment similar to the stable warped discs.

Misalignments are perhaps more likely in younger discs that are dynamically interacting with their surroundings and have a higher multiplicity, in which case the disc mass may be higher than the commonly assumed $\sim1$~per cent of the stellar mass. With a greater $M_{\rm{disc}}/M_{\rm{star}}$, the binary orbital evolution will be non-negligible and the binary and disc will align faster. More massive discs are likely to be self-gravitating but the additional heating generated by the shearing motions in a warped disc may act to stablize an otherwise gravitationally unstable disc \citep[see][]{rowther2022}. Further study is needed regarding the evolution of warps in young, more massive discs.

\subsection{Implications for observations}
\label{sec:implications}

The kinds of structures we expect from these models includes those already observed such as a warped inner region of the disc, discs with distinct misaligned inner and outer components and spirals. {\edit We also see more exotic structures such as near-polar rings (a feature that has been firmly identified in one disc to date; \citealt{kennedy2019,zuniga2021}) and messy unstable rings that fragment and reform.} The latter may also manifest as variations in the luminosity since the accretion rate is highly variable. Another interesting feature to highlight is the ring of reduced surface density in the discs with a large warp that fail to break (see Fig.~\ref{fig:misalignment_renders}, top left panel). This may be observed as a gap, with no obvious formation mechanism. There is evidence from simulations that dust rings form as dust piles up at a warp maximum independently of any dust trapping due to pressure bumps \citep{aly2020}, which would instead lead to a ring forming. This is a topic that needs to be explored further: should we expect to see a ring or a gap in submillimetre dust emission coincident with a warp? {\edit Simulations show that long-lived dust rings may form in unbroken warped discs \citep{aly2020,aly2021} and these might give the appearance of a broken disc. When such structures are observed, it is therefore necessary to check whether the gas component displays a break as well. }

Several discs have been observed to have a separate misaligned inner disc as opposed to a smooth warp. While \citet{facchini2014} showed that even a massive planetary companion could warp a disc, the simulations presented here imply that only {\edit binaries with $\mu>0.1$} can tear their disc, which effectively precludes the possibility of an unseen stellar or planetary companion causing a misaligned inner disc {\edit from within the central cavity through this process}. {\edit Other scenarios may better explain the presence of misaligned inner discs where there is no companion within the cavity that fits the criteria. Other work indicates that an external binary can induce a small warp when the outer radius lies near a resonance \citep{lubow2000}. However, a misaligned circumstellar disc tends to align much more quickly than a circumbinary disc \citep[e.g.][]{foucart2014}. Alternatively, a planet with an orbit inclined with respect to the disc that has carved a gap in the disc may cause the disc to warp or cause the inner and outer components to precess independently, without the presence of a stellar companion \citep{xiang-gruess2013,nealon2018,zhu2019}.} Another explanation for misaligned inner and outer discs is that a planet has created a discontinuity (i.e. a gap in the surface density profile) in the disc which has enabled a disc misaligned to the orbit of a binary to break cleanly when it would otherwise have remained smoothly warped \citep[cf.][]{xiang-gruess2014}. These are different processes to disc tearing by a central (or external) companion since the disc is broken by the combination of the Lindblad torques exerted by the planet and the precession torques.

In section \ref{sec:discussion_factors} we pointed out that a disc torn by a central binary exhibits a nearly planar outer disc with only a modest warp. Therefore, warps and tears in circumbinary discs may be somewhat mutually exclusive. If a broken disc is observed to have a strongly warped outer disc, this might indicate that a different mechanism is responsible for the break. A disc broken by a planet on a misaligned orbit should be warped beyond the gap carved by the planet \citep{xiang-gruess2013,nealon2018,nealon2019aa,zhu2019} and a disc warped by an \textit{external} companion may also be warped to the edge.

\subsection{Implications for the interpretation of GW Orionis}

{\edit Observations of the GW Orionis triple system, presented by \cite{kraus2020}, provide the most compelling evidence for disc tearing driven by gravitational torques from the stellar components.} The GW Orionis disc displays clear rings in submillimetre emission along with the twisted velocity map characteristic of a warp. The innermost circumtriple ring casts a shadow and therefore is interpreted as a torn inner disc that is highly misaligned and eccentric. There are discrepancies between simulations of the GW Orionis disc which means it has not been possible to attribute the torn structure definitively to the precession torques of the triple star system or to a putative embedded planet \citep{kraus2020,smallwood2021}.

One of the issues raised is of whether a stellar companion can cause a disc to break at a radius of $\sim50$~au. In simulations we have obtained breaking radii of $\sim5.7a_{\rm b}$ - $11a_{\rm b}$ in terms of the semi-major axis of the secondary $a_{\rm b}$, which is greater than the simple theoretical arguments predict (see section \ref{sec:rbreak}). In terms of the semi-major axis of the tertiary component, the observed break radius in GW Orionis is $\sim 5.6 a_c$, which could plausibly be due to the stellar torques. The mass ratio is $M_{\rm c}/(M_{\rm a}+M_{\rm b}+M_{\rm c}) = 0.26$, the misalignment between outer disc and the orbit of the tertiary is $28\degree$, and the eccentricity is $e_{\rm c}\approx0.4$. With these parameters, and the results presented here, the disc would be expected to have a steep warp and be borderline unstable. The disc is unusually massive and this would increase the precession torque, potentially reducing the stability of the disc further. The inner binary may be able to tear a disc, since $M_{\rm b}/(M_{\rm a}+ M_{\rm b}) = 0.37$ and the misalignment is $\approx 42\degree$. However $a_{\rm b} = 1.2$~au and the disc is truncated outside the orbit of the tertiary at $R>10$~au, where the torque from the central binary is much reduced.

The simulations of \citet{smallwood2021} show that the GW Orionis disc can break (their Run 10). The authors discount this result because breaking was observed only with the initial $R_{\rm in} =20$~au and not with $R_{\rm in} =30$~au. Nevertheless, we can estimate that the disc should be truncated at $\sim 2.5 a = 22$~au (see \citealt{miranda2015} and section \ref{sec:inneredge}) and, indeed, CO emission is detected within the inner circumtriple ring \citep{kraus2020}. The dust emission from close to the stars indicates that material is transported all the way to the central cavity since this would need frequent replenishment. An inner radius $<30$~au is therefore quite likely.

A final point is that there is an uncertainty of several degrees in the misalignment between stellar orbits and the disc due to the difficulty of deriving the disc inclination and position angle. Furthermore, the quoted uncertainties of the stellar masses allow a mass ratio up to $M_{\rm c}/(M_{\rm a}+M_{\rm b}+M_{\rm c}) = 0.32$ (using the orbits derived in \citealt{kraus2020}). This highlights the need to test the full range of parameters allowed by observations because the difference in parameters due to observational uncertainties can produce discs with very different structures. While it is clear there is much we do not understand about this system, the results of our simulations suggest that disc tearing caused by the precession torques from the triple could be responsible for the production of the ring \citep[as argued by][]{kraus2020}.

\subsection{Lessons for future modelling}
\label{sec:lessons}

Hydrodynamical models are often invoked to explain the structures of the many discs now found to have evidence of misalignments. Having studied a wide range of parameters, we now summarise some general points for consideration when modelling warped and torn discs.

A well-known difficulty of constructing a model of a particular disc is that of achieving sufficient resolution near the centre, while modelling the disc to a large enough radius to produce an accurate result. This problem can be avoided somewhat in grid-based codes where a logarithmically spaced radial grid can be implemented, but this is not an option for SPH codes. Can we make do with modelling just the inner regions of a disc and recover the correct structure? In section \ref{sec:r_out} we showed that the structure of the inner 20~au of a 50~au disc is very similar when $R_{\rm{out}}=25$~au. It is therefore acceptable to model the inner region of a larger disc as long as the simulation time is shorter than the time taken for the warp to reach the edge of the smaller disc. The tilt and twist of the outer edge of the smaller disc is not representative of the full disc and should be ignored. For studying the longer term evolution of a disc for $t_{\rm{max}}> t_{\rm{comm}}$, the whole disc will need to be modelled.

The choice of reasonable initial conditions is a broad issue for disc modelling in general but especially for the discs with warps and misalignments, which are not in a steady state. Late accretion of material from the surrounding cloud is perhaps most likely to be the origin of misaligned discs, in which case we cannot expect the disc to have a well-defined outer radius. The SPH models of \citet{hirsh2020} showed that the size of the central cavity can take tens of thousands of binary orbits to settle fully. In nature, the inner edge will not necessarily have settled because the accretion rate will be variable on shorter timescales and the disc inclination will be readjusting. 
The behaviour of a misaligned disc is sensitive to the inner edge location and the resolution there. Therefore, a good approach for modelling an observed disc is to test different values of $R_{\rm{in}}$ and compare the evolution. Additionally, we emphasize that it is the extent of the gas disc that matters whereas the inner and outer edges are often derived from observations of dust emission.

A further consideration is that the variations in the values measured for disc properties are considerable, often greater than the quoted uncertainties. Often, the range of possible values for some parameters would produce discs with different structures in simulations. We need to be careful about drawing conclusions from simulations of observed discs and should ensure that the whole spread of parameters allowed from observational constraints have been studied.

\section{Conclusion}

We have performed a number of simulations to better understand the structures formed in misaligned protoplanetary discs around binaries. These simulations confirm that disc tearing can occur in circumbinary protoplanetary discs, even with a relatively thick disc with $H/R \geq 0.05$. In general, the structure of the disc depends on the combination of properties, rather than on any single parameter. Our simulations with a low binary mass ratio {\edit ($\mu =0.1$) do not exhibit disc tearing}, which implies that a planetary companion  is unlikely to tear a disc from within the central cavity. The misalignment angle required for tearing depends on the binary mass ratio and only discs with misalignment $\gtrsim40\degree$ showed disc tearing in our simulations. Highly eccentric binaries are an exception to this and can tear their discs with a smaller misalignment angle. A disc misaligned by just $25\degree$ can be torn if the binary eccentricity $e\gtrsim 0.8$. The surface density and sound speed profiles can also play a role in determining the stability of the disc and should not be ignored in marginally stable cases.

In protoplanetary circumbinary discs, the warp is driven almost entirely from the inner regions of the disc because the precession torque decreases sharply with radius and bending waves communicate the warp efficiently to larger radii. This is different to discs in which the warp propagates via diffusion, where the structure is determined by the local precession torque in the disc. Therefore, the location of the inner edge is an important parameter for determining the evolution of a misaligned protoplanetary disc since it directly affects the strength of the precession torque that drives the warp which shapes the whole disc. For protoplanetary discs that exhibit a break caused by the precession from the binary, the outer disc is typically largely unwarped. This suggests that protoplanetary discs with strongly warped outer discs are sculpted by an external companion or an embedded planet rather than an inner binary.

At present, we cannot {\it a priori} link the radius at which the disc breaks to the initial properties of the binary/disc. In all cases, the breaking radius was significantly greater than predicted from simple theoretical arguments and the location of the break is not constant in time. {\edit The analytical arguments for breaking wavelike discs \citep[see the appendix of][]{nixon2013} assume that the warp at a given radius is driven solely by the precession torque at that radius, but as discussed above, this is not true for wavelike discs where the warp at any radius can be driven by a combination of local precession and propagation of warps from smaller (or larger) radii.} It is also likely that localised dissipative effects come into play and this warrants further investigation. Moreover, real misaligned protoplanetary discs are not in a steady state because of recent/ongoing accretion that may be the origin of the misalignment and/or the disc and binary are evolving towards alignment. This creates difficulties for applying simple criteria directly to observed systems.

In addition to existing observational signatures, {\edit we note that the ring of reduced surface density that is formed when a disc is significantly warped may generate observable consequences}. This might be linked to a dust gap but it is not clear how this would appear if the dust is also shaped by precession torques independently of the gas {\edit as suggested by \citep{aly2020}}. {\edit While it is clear from our results that determining where and when a wavelike disc can break is not a trivial task, it is also clear that by combining sophisticated simulation models with detailed observations of spatially resolved protoplanetary discs we can make significant progress in linking observed structures with the physical mechanisms driving them.}

\section*{Acknowledgements}
We thank the reviewer for their comments and suggestions. This research made use of the DiRAC Data Intensive service at Leicester, operated by the University of Leicester IT Services, which forms part of the Science and Technology Facilities Council (STFC) DiRAC HPC Facility (\url{www.dirac.ac.uk}). The equipment was funded by BEIS capital funding via STFC capital grants ST/K000373/1 and ST/R002363/1 and STFC DiRAC Operations grant ST/R001014/1. DiRAC is part of the National e-Infrastructure. AKY and KR are grateful for support from the UK STFC via grant ST/V000594/1. SS was supported by a University of Edinburgh School of Physics and Astronomy Career Development Summer Scholarship. CJN acknowledges support from the Science and Technology Facilities Council (grant number ST/Y000544/1), and the Leverhulme Trust (grant number RPG-2021-380). This work made use of {\sc numpy} \citep{harris2020}, {\sc matplotlib} \citep{hunter2007} and {\sc splash} \citep{price2007}.
\section*{Data Availability}

The simulations were performed with {\sc phantom} which is available from \url{https://github.com/danieljprice/phantom}. Files for generating the models in this paper and the analysis files for producing the plots will be made available online via Edinburgh Datashare.


\typeout{} 
\bibliographystyle{mnras}
\bibliography{papers} 

\begin{thebibliography}{}
\makeatletter
\relax
\def\mn@urlcharsother{\let\do\@makeother \do\$\do\&\do\#\do\^\do\_\do\%\do\~}
\def\mn@doi{\begingroup\mn@urlcharsother \@ifnextchar [ {\mn@doi@}
  {\mn@doi@[]}}
\def\mn@doi@[#1]#2{\def\@tempa{#1}\ifx\@tempa\@empty \href
  {http://dx.doi.org/#2} {doi:#2}\else \href {http://dx.doi.org/#2} {#1}\fi
  \endgroup}
\def\mn@eprint#1#2{\mn@eprint@#1:#2::\@nil}
\def\mn@eprint@arXiv#1{\href {http://arxiv.org/abs/#1} {{\tt arXiv:#1}}}
\def\mn@eprint@dblp#1{\href {http://dblp.uni-trier.de/rec/bibtex/#1.xml}
  {dblp:#1}}
\def\mn@eprint@#1:#2:#3:#4\@nil{\def\@tempa {#1}\def\@tempb {#2}\def\@tempc
  {#3}\ifx \@tempc \@empty \let \@tempc \@tempb \let \@tempb \@tempa \fi \ifx
  \@tempb \@empty \def\@tempb {arXiv}\fi \@ifundefined
  {mn@eprint@\@tempb}{\@tempb:\@tempc}{\expandafter \expandafter \csname
  mn@eprint@\@tempb\endcsname \expandafter{\@tempc}}}

\bibitem[\protect\citeauthoryear{{Aly} \& {Lodato}}{{Aly} \&
  {Lodato}}{2020}]{aly2020}
{Aly} H.,  {Lodato} G.,  2020, \mn@doi [\mnras] {10.1093/mnras/stz3633}, \href
  {https://ui.adsabs.harvard.edu/abs/2020MNRAS.tmp.3252A} {p.~3252}

\bibitem[\protect\citeauthoryear{{Aly}, {Dehnen}, {Nixon}  \& {King}}{{Aly}
  et~al.}{2015}]{aly2015}
{Aly} H.,  {Dehnen} W.,  {Nixon} C.,   {King} A.,  2015, \mn@doi [\mnras]
  {10.1093/mnras/stv128}, \href
  {https://ui.adsabs.harvard.edu/abs/2015MNRAS.449...65A} {449, 65}

\bibitem[\protect\citeauthoryear{{Aly}, {Gonzalez}, {Nealon}, {Longarini},
  {Lodato}  \& {Price}}{{Aly} et~al.}{2021}]{aly2021}
{Aly} H.,  {Gonzalez} J.-F.,  {Nealon} R.,  {Longarini} C.,  {Lodato} G.,
  {Price} D.~J.,  2021, \mn@doi [\mnras] {10.1093/mnras/stab2794}, \href
  {https://ui.adsabs.harvard.edu/abs/2021MNRAS.508.2743A} {508, 2743}

\bibitem[\protect\citeauthoryear{{Ansdell} et~al.,}{{Ansdell}
  et~al.}{2020}]{ansdell2020}
{Ansdell} M.,  et~al., 2020, \mn@doi [\mnras] {10.1093/mnras/stz3361}, \href
  {https://ui.adsabs.harvard.edu/abs/2020MNRAS.492..572A} {492, 572}

\bibitem[\protect\citeauthoryear{{Artymowicz} \& {Lubow}}{{Artymowicz} \&
  {Lubow}}{1994}]{artymowics1994}
{Artymowicz} P.,  {Lubow} S.~H.,  1994, \mn@doi [\apj] {10.1086/173679}, \href
  {https://ui.adsabs.harvard.edu/abs/1994ApJ...421..651A} {421, 651}

\bibitem[\protect\citeauthoryear{{Ballabio}, {Nealon}, {Alexander}, {Cuello},
  {Pinte}  \& {Price}}{{Ballabio} et~al.}{2021}]{ballabio2021}
{Ballabio} G.,  {Nealon} R.,  {Alexander} R.~D.,  {Cuello} N.,  {Pinte} C.,
  {Price} D.~J.,  2021, \mn@doi [\mnras] {10.1093/mnras/stab922}, \href
  {https://ui.adsabs.harvard.edu/abs/2021MNRAS.504..888B} {504, 888}

\bibitem[\protect\citeauthoryear{{Bate}, {Bonnell}  \& {Price}}{{Bate}
  et~al.}{1995}]{bate1995aa}
{Bate} M.~R.,  {Bonnell} I.~A.,   {Price} N.~M.,  1995, \mn@doi [\mnras]
  {10.1093/mnras/277.2.362}, \href
  {https://ui.adsabs.harvard.edu/#abs/1995MNRAS.277..362B} {277, 362}

\bibitem[\protect\citeauthoryear{{Benisty} et~al.,}{{Benisty}
  et~al.}{2017}]{benisty2017}
{Benisty} M.,  et~al., 2017, \mn@doi [\aap] {10.1051/0004-6361/201629798},
  \href {https://ui.adsabs.harvard.edu/abs/2017A&A...597A..42B} {597, A42}

\bibitem[\protect\citeauthoryear{{Bi} et~al.,}{{Bi} et~al.}{2020}]{bi2020}
{Bi} J.,  et~al., 2020, \mn@doi [\apjl] {10.3847/2041-8213/ab8eb4}, \href
  {https://ui.adsabs.harvard.edu/abs/2020ApJ...895L..18B} {895, L18}

\bibitem[\protect\citeauthoryear{{Casassus}, {P{\'e}rez}, {Osses}  \&
  {Marino}}{{Casassus} et~al.}{2019}]{casassus2019}
{Casassus} S.,  {P{\'e}rez} S.,  {Osses} A.,   {Marino} S.,  2019, \mn@doi
  [\mnras] {10.1093/mnrasl/slz059}, \href
  {https://ui.adsabs.harvard.edu/abs/2019MNRAS.486L..58C} {486, L58}

\bibitem[\protect\citeauthoryear{{Chiang} \& {Goldreich}}{{Chiang} \&
  {Goldreich}}{1997}]{chiang1997}
{Chiang} E.~I.,  {Goldreich} P.,  1997, \apj, 490, 368

\bibitem[\protect\citeauthoryear{{Cullen} \& {Dehnen}}{{Cullen} \&
  {Dehnen}}{2010}]{cullen2010}
{Cullen} L.,  {Dehnen} W.,  2010, \mn@doi [\mnras]
  {10.1111/j.1365-2966.2010.17158.x}, \href
  {https://ui.adsabs.harvard.edu/abs/2010MNRAS.408..669C} {408, 669}

\bibitem[\protect\citeauthoryear{{Czekala}, {Chiang}, {Andrews}, {Jensen},
  {Torres}, {Wilner}, {Stassun}  \& {Macintosh}}{{Czekala}
  et~al.}{2019}]{czekala2019}
{Czekala} I.,  {Chiang} E.,  {Andrews} S.~M.,  {Jensen} E. L.~N.,  {Torres} G.,
   {Wilner} D.~J.,  {Stassun} K.~G.,   {Macintosh} B.,  2019, \mn@doi [\apj]
  {10.3847/1538-4357/ab287b}, \href
  {https://ui.adsabs.harvard.edu/abs/2019ApJ...883...22C} {883, 22}

\bibitem[\protect\citeauthoryear{{D'Alessio}, {Cant{\"o}}, {Calvet}  \&
  {Lizano}}{{D'Alessio} et~al.}{1998}]{dalessio1998}
{D'Alessio} P.,  {Cant{\"o}} J.,  {Calvet} N.,   {Lizano} S.,  1998, \apj, 500,
  411

\bibitem[\protect\citeauthoryear{{Debes} et~al.,}{{Debes}
  et~al.}{2017}]{debes2017}
{Debes} J.~H.,  et~al., 2017, \mn@doi [\apj] {10.3847/1538-4357/835/2/205},
  \href {https://ui.adsabs.harvard.edu/abs/2017ApJ...835..205D} {835, 205}

\bibitem[\protect\citeauthoryear{{Deng} \& {Ogilvie}}{{Deng} \&
  {Ogilvie}}{2022}]{deng2022}
{Deng} H.,  {Ogilvie} G.~I.,  2022, \mn@doi [\mnras] {10.1093/mnras/stac858},
  \href {https://ui.adsabs.harvard.edu/abs/2022MNRAS.512.6078D} {512, 6078}

\bibitem[\protect\citeauthoryear{{Do{\u{g}}an} \& {Nixon}}{{Do{\u{g}}an} \&
  {Nixon}}{2020}]{dogan2020}
{Do{\u{g}}an} S.,  {Nixon} C.~J.,  2020, \mn@doi [\mnras]
  {10.1093/mnras/staa1239}, \href
  {https://ui.adsabs.harvard.edu/abs/2020MNRAS.495.1148D} {495, 1148}

\bibitem[\protect\citeauthoryear{{Do{\u{g}}an}, {Nixon}, {King}  \&
  {Price}}{{Do{\u{g}}an} et~al.}{2015}]{dogan2015}
{Do{\u{g}}an} S.,  {Nixon} C.,  {King} A.,   {Price} D.~J.,  2015, \mn@doi
  [\mnras] {10.1093/mnras/stv347}, \href
  {https://ui.adsabs.harvard.edu/abs/2015MNRAS.449.1251D} {449, 1251}

\bibitem[\protect\citeauthoryear{{Do{\v{g}}an}, {Nixon}, {King}  \&
  {Pringle}}{{Do{\v{g}}an} et~al.}{2018}]{dogan2018}
{Do{\v{g}}an} S.,  {Nixon} C.~J.,  {King} A.~R.,   {Pringle} J.~E.,  2018,
  \mn@doi [\mnras] {10.1093/mnras/sty155}, \href
  {https://ui.adsabs.harvard.edu/abs/2018MNRAS.476.1519D} {476, 1519}

\bibitem[\protect\citeauthoryear{{Drewes} \& {Nixon}}{{Drewes} \&
  {Nixon}}{2021}]{drewes2021}
{Drewes} N.~C.,  {Nixon} C.~J.,  2021, \mn@doi [\apj]
  {10.3847/1538-4357/ac2609}, \href
  {https://ui.adsabs.harvard.edu/abs/2021ApJ...922..243D} {922, 243}

\bibitem[\protect\citeauthoryear{{Facchini}, {Lodato}  \& {Price}}{{Facchini}
  et~al.}{2013}]{facchini2013}
{Facchini} S.,  {Lodato} G.,   {Price} D.~J.,  2013, \mn@doi [\mnras]
  {10.1093/mnras/stt877}, \href
  {https://ui.adsabs.harvard.edu/abs/2013MNRAS.433.2142F} {433, 2142}

\bibitem[\protect\citeauthoryear{{Facchini}, {Ricci}  \& {Lodato}}{{Facchini}
  et~al.}{2014}]{facchini2014}
{Facchini} S.,  {Ricci} L.,   {Lodato} G.,  2014, \mn@doi [\mnras]
  {10.1093/mnras/stu1149}, \href
  {https://ui.adsabs.harvard.edu/abs/2014MNRAS.442.3700F} {442, 3700}

\bibitem[\protect\citeauthoryear{{Facchini}, {Juh{\'a}sz}  \&
  {Lodato}}{{Facchini} et~al.}{2018}]{facchini2018}
{Facchini} S.,  {Juh{\'a}sz} A.,   {Lodato} G.,  2018, \mn@doi [\mnras]
  {10.1093/mnras/stx2523}, \href
  {https://ui.adsabs.harvard.edu/abs/2018MNRAS.473.4459F} {473, 4459}

\bibitem[\protect\citeauthoryear{{Fang}, {Sicilia-Aguilar}, {Roccatagliata},
  {Fedele}, {Henning}, {Eiroa}  \& {M{\"u}ller}}{{Fang}
  et~al.}{2014}]{fang2014}
{Fang} M.,  {Sicilia-Aguilar} A.,  {Roccatagliata} V.,  {Fedele} D.,  {Henning}
  T.,  {Eiroa} C.,   {M{\"u}ller} A.,  2014, \mn@doi [\aap]
  {10.1051/0004-6361/201424146}, \href
  {https://ui.adsabs.harvard.edu/abs/2014A&A...570A.118F} {570, A118}

\bibitem[\protect\citeauthoryear{{Foucart} \& {Lai}}{{Foucart} \&
  {Lai}}{2014}]{foucart2014}
{Foucart} F.,  {Lai} D.,  2014, \mn@doi [\mnras] {10.1093/mnras/stu1869}, \href
  {https://ui.adsabs.harvard.edu/abs/2014MNRAS.445.1731F} {445, 1731}

\bibitem[\protect\citeauthoryear{{Fragner} \& {Nelson}}{{Fragner} \&
  {Nelson}}{2010}]{fragner2010}
{Fragner} M.~M.,  {Nelson} R.~P.,  2010, \mn@doi [\aap]
  {10.1051/0004-6361/200913088}, \href
  {https://ui.adsabs.harvard.edu/abs/2010A&A...511A..77F} {511, A77}

\bibitem[\protect\citeauthoryear{{Gammie} \& {Menou}}{{Gammie} \&
  {Menou}}{1998}]{gammie1998}
{Gammie} C.~F.,  {Menou} K.,  1998, \mn@doi [\apjl] {10.1086/311091}, \href
  {https://ui.adsabs.harvard.edu/abs/1998ApJ...492L..75G} {492, L75}

\bibitem[\protect\citeauthoryear{Harris et~al.,}{Harris
  et~al.}{2020}]{harris2020}
Harris C.~R.,  et~al., 2020, Nature, 585, 357

\bibitem[\protect\citeauthoryear{{Hartmann}, {Calvet}, {Gullbring}  \&
  {D'Alessio}}{{Hartmann} et~al.}{1998}]{hartmann1998}
{Hartmann} L.,  {Calvet} N.,  {Gullbring} E.,   {D'Alessio} P.,  1998, \mn@doi
  [\apj] {10.1086/305277}, \href
  {https://ui.adsabs.harvard.edu/abs/1998ApJ...495..385H} {495, 385}

\bibitem[\protect\citeauthoryear{{Heath} \& {Nixon}}{{Heath} \&
  {Nixon}}{2020}]{heath2020}
{Heath} R.~M.,  {Nixon} C.~J.,  2020, \mn@doi [\aap]
  {10.1051/0004-6361/202038548}, \href
  {https://ui.adsabs.harvard.edu/abs/2020A&A...641A..64H} {641, A64}

\bibitem[\protect\citeauthoryear{{Hirsh}, {Price}, {Gonzalez},
  {Ubeira-Gabellini}  \& {Ragusa}}{{Hirsh} et~al.}{2020}]{hirsh2020}
{Hirsh} K.,  {Price} D.~J.,  {Gonzalez} J.-F.,  {Ubeira-Gabellini} M.~G.,
  {Ragusa} E.,  2020, \mn@doi [\mnras] {10.1093/mnras/staa2536}, \href
  {https://ui.adsabs.harvard.edu/abs/2020MNRAS.498.2936H} {498, 2936}

\bibitem[\protect\citeauthoryear{{Hunter, J. D.}}{{Hunter, J.
  D.}}{2007}]{hunter2007}
{Hunter, J. D.} 2007, {Computing in Science \& Engineering}, 9, 90

\bibitem[\protect\citeauthoryear{{Juh{\'a}sz} \& {Facchini}}{{Juh{\'a}sz} \&
  {Facchini}}{2017}]{juhasz2017}
{Juh{\'a}sz} A.,  {Facchini} S.,  2017, \mn@doi [\mnras]
  {10.1093/mnras/stw3389}, \href
  {https://ui.adsabs.harvard.edu/abs/2017MNRAS.466.4053J} {466, 4053}

\bibitem[\protect\citeauthoryear{{Kennedy} et~al.,}{{Kennedy}
  et~al.}{2019}]{kennedy2019}
{Kennedy} G.~M.,  et~al., 2019, \mn@doi [Nature Astronomy]
  {10.1038/s41550-018-0667-x}, \href
  {https://ui.adsabs.harvard.edu/abs/2019NatAs...3..230K} {3, 230}

\bibitem[\protect\citeauthoryear{{Keyte} et~al.,}{{Keyte}
  et~al.}{2023}]{keyte2023}
{Keyte} L.,  et~al., 2023, \mn@doi [arXiv e-prints]
  {10.48550/arXiv.2303.08927}, \href
  {https://ui.adsabs.harvard.edu/abs/2023arXiv230308927K} {p. arXiv:2303.08927}

\bibitem[\protect\citeauthoryear{{Kraus} et~al.,}{{Kraus}
  et~al.}{2020}]{kraus2020}
{Kraus} S.,  et~al., 2020, \mn@doi [Science] {10.1126/science.aba4633}, \href
  {https://ui.adsabs.harvard.edu/abs/2020Sci...369.1233K} {369, 1233}

\bibitem[\protect\citeauthoryear{{Lakeland} \& {Naylor}}{{Lakeland} \&
  {Naylor}}{2022}]{lakeland2022}
{Lakeland} B.~S.,  {Naylor} T.,  2022, \mn@doi [\mnras]
  {10.1093/mnras/stac1477}, \href
  {https://ui.adsabs.harvard.edu/abs/2022MNRAS.514.2736L} {514, 2736}

\bibitem[\protect\citeauthoryear{{Larwood} \& {Papaloizou}}{{Larwood} \&
  {Papaloizou}}{1997}]{larwood1997}
{Larwood} J.~D.,  {Papaloizou} J. C.~B.,  1997, \mn@doi [\mnras]
  {10.1093/mnras/285.2.288}, \href
  {https://ui.adsabs.harvard.edu/abs/1997MNRAS.285..288L} {285, 288}

\bibitem[\protect\citeauthoryear{{Larwood}, {Nelson}, {Papaloizou}  \&
  {Terquem}}{{Larwood} et~al.}{1996}]{larwood1996}
{Larwood} J.~D.,  {Nelson} R.~P.,  {Papaloizou} J.~C.~B.,   {Terquem} C.,
  1996, \mn@doi [\mnras] {10.1093/mnras/282.2.597}, \href
  {https://ui.adsabs.harvard.edu/abs/1996MNRAS.282..597L} {282, 597}

\bibitem[\protect\citeauthoryear{{Laws} et~al.,}{{Laws}
  et~al.}{2020}]{laws2020}
{Laws} A. S.~E.,  et~al., 2020, \mn@doi [\apj] {10.3847/1538-4357/ab59e2},
  \href {https://ui.adsabs.harvard.edu/abs/2020ApJ...888....7L} {888, 7}

\bibitem[\protect\citeauthoryear{{Liska}, {Hesp}, {Tchekhovskoy}, {Ingram},
  {van der Klis}, {Markoff}  \& {Van Moer}}{{Liska} et~al.}{2021}]{liska2021}
{Liska} M.,  {Hesp} C.,  {Tchekhovskoy} A.,  {Ingram} A.,  {van der Klis} M.,
  {Markoff} S.~B.,   {Van Moer} M.,  2021, \mnras, 507, 983

\bibitem[\protect\citeauthoryear{{Lodato} \& {Price}}{{Lodato} \&
  {Price}}{2010}]{lodato2010}
{Lodato} G.,  {Price} D.~J.,  2010, \mn@doi [\mnras]
  {10.1111/j.1365-2966.2010.16526.x}, \href
  {https://ui.adsabs.harvard.edu/abs/2010MNRAS.405.1212L} {405, 1212}

\bibitem[\protect\citeauthoryear{{Lodato} \& {Pringle}}{{Lodato} \&
  {Pringle}}{2006}]{lodato2006}
{Lodato} G.,  {Pringle} J.~E.,  2006, \mn@doi [\mnras]
  {10.1111/j.1365-2966.2006.10194.x}, \href
  {https://ui.adsabs.harvard.edu/abs/2006MNRAS.368.1196L} {368, 1196}

\bibitem[\protect\citeauthoryear{{Lubow} \& {Ogilvie}}{{Lubow} \&
  {Ogilvie}}{2000}]{lubow2000}
{Lubow} S.~H.,  {Ogilvie} G.~I.,  2000, \mn@doi [\apj] {10.1086/309101}, \href
  {https://ui.adsabs.harvard.edu/abs/2000ApJ...538..326L} {538, 326}

\bibitem[\protect\citeauthoryear{{Lubow}, {Ogilvie}  \& {Pringle}}{{Lubow}
  et~al.}{2002}]{lubow2002}
{Lubow} S.~H.,  {Ogilvie} G.~I.,   {Pringle} J.~E.,  2002, \mn@doi [\mnras]
  {10.1046/j.1365-8711.2002.05949.x}, \href
  {https://ui.adsabs.harvard.edu/abs/2002MNRAS.337..706L} {337, 706}

\bibitem[\protect\citeauthoryear{{Lubow}, {Martin}  \& {Nixon}}{{Lubow}
  et~al.}{2015}]{lubow2015}
{Lubow} S.~H.,  {Martin} R.~G.,   {Nixon} C.,  2015, \mn@doi [\apj]
  {10.1088/0004-637X/800/2/96}, \href
  {https://ui.adsabs.harvard.edu/abs/2015ApJ...800...96L} {800, 96}

\bibitem[\protect\citeauthoryear{{Marino}, {Perez}  \& {Casassus}}{{Marino}
  et~al.}{2015}]{marino2015}
{Marino} S.,  {Perez} S.,   {Casassus} S.,  2015, \mn@doi [\apjl]
  {10.1088/2041-8205/798/2/L44}, \href
  {https://ui.adsabs.harvard.edu/abs/2015ApJ...798L..44M} {798, L44}

\bibitem[\protect\citeauthoryear{{Martin} \& {Lubow}}{{Martin} \&
  {Lubow}}{2017}]{martin2017}
{Martin} R.~G.,  {Lubow} S.~H.,  2017, \mn@doi [\apjl]
  {10.3847/2041-8213/835/2/L28}, \href
  {https://ui.adsabs.harvard.edu/abs/2017ApJ...835L..28M} {835, L28}

\bibitem[\protect\citeauthoryear{{Martin}, {Nixon}, {Pringle}  \&
  {Livio}}{{Martin} et~al.}{2019}]{martin2019}
{Martin} R.~G.,  {Nixon} C.~J.,  {Pringle} J.~E.,   {Livio} M.,  2019, \mn@doi
  [\na] {10.1016/j.newast.2019.01.001}, \href
  {https://ui.adsabs.harvard.edu/abs/2019NewA...70....7M} {70, 7}

\bibitem[\protect\citeauthoryear{{Meru} \& {Bate}}{{Meru} \&
  {Bate}}{2012}]{meru2012}
{Meru} F.,  {Bate} M.~R.,  2012, \mn@doi [\mnras]
  {10.1111/j.1365-2966.2012.22035.x}, \href
  {https://ui.adsabs.harvard.edu/abs/2012MNRAS.427.2022M} {427, 2022}

\bibitem[\protect\citeauthoryear{{Min}, {Stolker}, {Dominik}  \&
  {Benisty}}{{Min} et~al.}{2017}]{min2017}
{Min} M.,  {Stolker} T.,  {Dominik} C.,   {Benisty} M.,  2017, \mn@doi [\aap]
  {10.1051/0004-6361/201730949}, \href
  {https://ui.adsabs.harvard.edu/abs/2017A&A...604L..10M} {604, L10}

\bibitem[\protect\citeauthoryear{{Miranda} \& {Lai}}{{Miranda} \&
  {Lai}}{2015}]{miranda2015}
{Miranda} R.,  {Lai} D.,  2015, \mn@doi [\mnras] {10.1093/mnras/stv1450}, \href
  {https://ui.adsabs.harvard.edu/abs/2015MNRAS.452.2396M} {452, 2396}

\bibitem[\protect\citeauthoryear{{Muro-Arena} et~al.,}{{Muro-Arena}
  et~al.}{2020}]{muro-arena2020}
{Muro-Arena} G.~A.,  et~al., 2020, \mn@doi [\aap]
  {10.1051/0004-6361/201936509}, \href
  {https://ui.adsabs.harvard.edu/abs/2020A&A...635A.121M} {635, A121}

\bibitem[\protect\citeauthoryear{{Nealon}, {Price}  \& {Nixon}}{{Nealon}
  et~al.}{2015}]{nealon2015}
{Nealon} R.,  {Price} D.~J.,   {Nixon} C.~J.,  2015, \mn@doi [\mnras]
  {10.1093/mnras/stv014}, \href
  {https://ui.adsabs.harvard.edu/abs/2015MNRAS.448.1526N} {448, 1526}

\bibitem[\protect\citeauthoryear{{Nealon}, {Dipierro}, {Alexander}, {Martin}
  \& {Nixon}}{{Nealon} et~al.}{2018}]{nealon2018}
{Nealon} R.,  {Dipierro} G.,  {Alexander} R.,  {Martin} R.~G.,   {Nixon} C.,
  2018, \mn@doi [\mnras] {10.1093/mnras/sty2267}, \href
  {https://ui.adsabs.harvard.edu/abs/2018MNRAS.481...20N} {481, 20}

\bibitem[\protect\citeauthoryear{{Nealon}, {Pinte}, {Alexander}, {Mentiplay}
  \& {Dipierro}}{{Nealon} et~al.}{2019}]{nealon2019aa}
{Nealon} R.,  {Pinte} C.,  {Alexander} R.,  {Mentiplay} D.,   {Dipierro} G.,
  2019, \mn@doi [\mnras] {10.1093/mnras/stz346}, \href
  {https://ui.adsabs.harvard.edu/abs/2019MNRAS.484.4951N} {484, 4951}

\bibitem[\protect\citeauthoryear{{Nealon}, {Ragusa}, {Gerosa}, {Rosotti}  \&
  {Barbieri}}{{Nealon} et~al.}{2022}]{nealon2022}
{Nealon} R.,  {Ragusa} E.,  {Gerosa} D.,  {Rosotti} G.,   {Barbieri} R.,  2022,
  \mnras, 509, 5608

\bibitem[\protect\citeauthoryear{{Nixon}}{{Nixon}}{2012}]{nixon2012jul}
{Nixon} C.~J.,  2012, \mn@doi [\mnras] {10.1111/j.1365-2966.2012.21072.x},
  \href {https://ui.adsabs.harvard.edu/abs/2012MNRAS.423.2597N} {423, 2597}

\bibitem[\protect\citeauthoryear{{Nixon} \& {King}}{{Nixon} \&
  {King}}{2012}]{nixon2012apr}
{Nixon} C.~J.,  {King} A.~R.,  2012, \mn@doi [\mnras]
  {10.1111/j.1365-2966.2011.20377.x}, \href
  {https://ui.adsabs.harvard.edu/abs/2012MNRAS.421.1201N} {421, 1201}

\bibitem[\protect\citeauthoryear{{Nixon} \& {Lubow}}{{Nixon} \&
  {Lubow}}{2015}]{nixonlubow2015}
{Nixon} C.,  {Lubow} S.~H.,  2015, \mn@doi [\mnras] {10.1093/mnras/stv166},
  \href {https://ui.adsabs.harvard.edu/abs/2015MNRAS.448.3472N} {448, 3472}

\bibitem[\protect\citeauthoryear{{Nixon} \& {Pringle}}{{Nixon} \&
  {Pringle}}{2010}]{nixon2010}
{Nixon} C.~J.,  {Pringle} J.~E.,  2010, \mn@doi [\mnras]
  {10.1111/j.1365-2966.2010.16331.x}, \href
  {https://ui.adsabs.harvard.edu/abs/2010MNRAS.403.1887N} {403, 1887}

\bibitem[\protect\citeauthoryear{{Nixon}, {Cossins}, {King}  \&
  {Pringle}}{{Nixon} et~al.}{2011}]{nixon2011}
{Nixon} C.~J.,  {Cossins} P.~J.,  {King} A.~R.,   {Pringle} J.~E.,  2011,
  \mn@doi [\mnras] {10.1111/j.1365-2966.2010.17952.x}, \href
  {https://ui.adsabs.harvard.edu/abs/2011MNRAS.412.1591N} {412, 1591}

\bibitem[\protect\citeauthoryear{{Nixon}, {King}, {Price}  \& {Frank}}{{Nixon}
  et~al.}{2012}]{nixon2012}
{Nixon} C.,  {King} A.,  {Price} D.,   {Frank} J.,  2012, \mn@doi [\apjl]
  {10.1088/2041-8205/757/2/L24}, \href
  {https://ui.adsabs.harvard.edu/abs/2012ApJ...757L..24N} {757, L24}

\bibitem[\protect\citeauthoryear{{Nixon}, {King}  \& {Price}}{{Nixon}
  et~al.}{2013}]{nixon2013}
{Nixon} C.,  {King} A.,   {Price} D.,  2013, \mn@doi [\mnras]
  {10.1093/mnras/stt1136}, \href
  {https://ui.adsabs.harvard.edu/abs/2013MNRAS.434.1946N} {434, 1946}

\bibitem[\protect\citeauthoryear{{Ogilvie}}{{Ogilvie}}{1999}]{ogilvie1999}
{Ogilvie} G.~I.,  1999, \mn@doi [\mnras] {10.1046/j.1365-8711.1999.02340.x},
  \href {https://ui.adsabs.harvard.edu/abs/1999MNRAS.304..557O} {304, 557}

\bibitem[\protect\citeauthoryear{{Ogilvie}}{{Ogilvie}}{2000}]{ogilvie2000}
{Ogilvie} G.~I.,  2000, \mn@doi [\mnras] {10.1046/j.1365-8711.2000.03654.x},
  \href {https://ui.adsabs.harvard.edu/abs/2000MNRAS.317..607O} {317, 607}

\bibitem[\protect\citeauthoryear{{Paneque-Carre{\~n}o}
  et~al.,}{{Paneque-Carre{\~n}o} et~al.}{2021}]{paneque2021}
{Paneque-Carre{\~n}o} T.,  et~al., 2021, \mn@doi [\apj]
  {10.3847/1538-4357/abf243}, \href
  {https://ui.adsabs.harvard.edu/abs/2021ApJ...914...88P} {914, 88}

\bibitem[\protect\citeauthoryear{{Papaloizou} \& {Pringle}}{{Papaloizou} \&
  {Pringle}}{1983}]{papaloizou1983}
{Papaloizou} J.~C.~B.,  {Pringle} J.~E.,  1983, \mn@doi [\mnras]
  {10.1093/mnras/202.4.1181}, \href
  {https://ui.adsabs.harvard.edu/abs/1983MNRAS.202.1181P} {202, 1181}

\bibitem[\protect\citeauthoryear{{Papaloizou} \& {Terquem}}{{Papaloizou} \&
  {Terquem}}{1995}]{papaloizou1995}
{Papaloizou} J. C.~B.,  {Terquem} C.,  1995, \mn@doi [\mnras]
  {10.1093/mnras/274.4.987}, \href
  {https://ui.adsabs.harvard.edu/abs/1995MNRAS.274..987P} {274, 987}

\bibitem[\protect\citeauthoryear{{P{\'e}rez} et~al.,}{{P{\'e}rez}
  et~al.}{2018}]{perez2018}
{P{\'e}rez} L.~M.,  et~al., 2018, \mn@doi [\apjl] {10.3847/2041-8213/aaf745},
  \href {https://ui.adsabs.harvard.edu/abs/2018ApJ...869L..50P} {869, L50}

\bibitem[\protect\citeauthoryear{{Petterson}}{{Petterson}}{1977}]{petterson1977}
{Petterson} J.~A.,  1977, \mn@doi [\apj] {10.1086/155280}, \href
  {https://ui.adsabs.harvard.edu/abs/1977ApJ...214..550P} {214, 550}

\bibitem[\protect\citeauthoryear{{Price}}{{Price}}{2007}]{price2007}
{Price} D.~J.,  2007, \mn@doi [\pasa] {10.1071/AS07022}, \href
  {http://adsabs.harvard.edu/abs/2007PASA...24..159P} {24, 159}

\bibitem[\protect\citeauthoryear{{Price} et~al.,}{{Price}
  et~al.}{2018}]{price2018aa}
{Price} D.~J.,  et~al., 2018, \mn@doi [Publications of the Astronomical Society
  of Australia] {10.1017/pasa.2018.25}, \href
  {https://ui.adsabs.harvard.edu/#abs/2018PASA...35...31P} {35, e031}

\bibitem[\protect\citeauthoryear{{Pringle}}{{Pringle}}{1992}]{pringle1992}
{Pringle} J.~E.,  1992, \mn@doi [\mnras] {10.1093/mnras/258.4.811}, \href
  {https://ui.adsabs.harvard.edu/abs/1992MNRAS.258..811P} {258, 811}

\bibitem[\protect\citeauthoryear{{Raj}, {Nixon}  \& {Do{\u{g}}an}}{{Raj}
  et~al.}{2021}]{raj2021a}
{Raj} A.,  {Nixon} C.~J.,   {Do{\u{g}}an} S.,  2021, \mn@doi [\apj]
  {10.3847/1538-4357/abdc24}, \href
  {https://ui.adsabs.harvard.edu/abs/2021ApJ...909...81R} {909, 81}

\bibitem[\protect\citeauthoryear{{Rosenfeld} et~al.,}{{Rosenfeld}
  et~al.}{2012}]{rosenfeld2012}
{Rosenfeld} K.~A.,  et~al., 2012, \mn@doi [\apj] {10.1088/0004-637X/757/2/129},
  \href {https://ui.adsabs.harvard.edu/abs/2012ApJ...757..129R} {757, 129}

\bibitem[\protect\citeauthoryear{{Rosotti}}{{Rosotti}}{2023}]{rosotti2023}
{Rosotti} G.~P.,  2023, \mn@doi [arXiv e-prints] {10.48550/arXiv.2302.01433},
  \href {https://ui.adsabs.harvard.edu/abs/2023arXiv230201433R} {p.
  arXiv:2302.01433}

\bibitem[\protect\citeauthoryear{{Rota} et~al.,}{{Rota}
  et~al.}{2022}]{rota2022}
{Rota} A.~A.,  et~al., 2022, \mn@doi [\aap] {10.1051/0004-6361/202141035},
  \href {https://ui.adsabs.harvard.edu/abs/2022A&A...662A.121R} {662, A121}

\bibitem[\protect\citeauthoryear{{Rowther}, {Nealon}  \& {Meru}}{{Rowther}
  et~al.}{2022}]{rowther2022}
{Rowther} S.,  {Nealon} R.,   {Meru} F.,  2022, \mn@doi [\apj]
  {10.3847/1538-4357/ac3975}, \href
  {https://ui.adsabs.harvard.edu/abs/2022ApJ...925..163R} {925, 163}

\bibitem[\protect\citeauthoryear{{Shakura} \& {Sunyaev}}{{Shakura} \&
  {Sunyaev}}{1973}]{shakura1973}
{Shakura} N. I.,  {Sunyaev} R.~A.,  1973, A\&A, 24, 337

\bibitem[\protect\citeauthoryear{{Smallwood}, {Nealon}, {Chen}, {Martin}, {Bi},
  {Dong}  \& {Pinte}}{{Smallwood} et~al.}{2021}]{smallwood2021}
{Smallwood} J.~L.,  {Nealon} R.,  {Chen} C.,  {Martin} R.~G.,  {Bi} J.,  {Dong}
  R.,   {Pinte} C.,  2021, \mn@doi [\mnras] {10.1093/mnras/stab2624}, \href
  {https://ui.adsabs.harvard.edu/abs/2021MNRAS.508..392S} {508, 392}

\bibitem[\protect\citeauthoryear{{Stevenson} \& {Young}}{{Stevenson} \&
  {Young}}{2022}]{stevenson2022}
{Stevenson} S.,  {Young} A.,  2022, \mn@doi [Research Notes of the American
  Astronomical Society] {10.3847/2515-5172/ac9afa}, \href
  {https://ui.adsabs.harvard.edu/abs/2022RNAAS...6..216S} {6, 216}

\bibitem[\protect\citeauthoryear{{Tremaine} \& {Davis}}{{Tremaine} \&
  {Davis}}{2014}]{tremaine2014}
{Tremaine} S.,  {Davis} S.~W.,  2014, \mn@doi [\mnras] {10.1093/mnras/stu663},
  \href {https://ui.adsabs.harvard.edu/abs/2014MNRAS.441.1408T} {441, 1408}

\bibitem[\protect\citeauthoryear{{Xiang-Gruess} \& {Papaloizou}}{{Xiang-Gruess}
  \& {Papaloizou}}{2013}]{xiang-gruess2013}
{Xiang-Gruess} M.,  {Papaloizou} J.~C.~B.,  2013, \mn@doi [\mnras]
  {10.1093/mnras/stt254}, \href
  {https://ui.adsabs.harvard.edu/abs/2013MNRAS.431.1320X} {431, 1320}

\bibitem[\protect\citeauthoryear{{Xiang-Gruess} \& {Papaloizou}}{{Xiang-Gruess}
  \& {Papaloizou}}{2014}]{xiang-gruess2014}
{Xiang-Gruess} M.,  {Papaloizou} J.~C.~B.,  2014, \mn@doi [\mnras]
  {10.1093/mnras/stu308}, \href
  {https://ui.adsabs.harvard.edu/abs/2014MNRAS.440.1179X} {440, 1179}

\bibitem[\protect\citeauthoryear{{Young}, {Alexander}, {Walsh}, {Nealon},
  {Booth}  \& {Pinte}}{{Young} et~al.}{2021}]{young2021}
{Young} A.~K.,  {Alexander} R.,  {Walsh} C.,  {Nealon} R.,  {Booth} A.,
  {Pinte} C.,  2021, \mn@doi [\mnras] {10.1093/mnras/stab1675}, \href
  {https://ui.adsabs.harvard.edu/abs/2021MNRAS.505.4821Y} {505, 4821}

\bibitem[\protect\citeauthoryear{{Young}, {Alexander}, {Rosotti}  \&
  {Pinte}}{{Young} et~al.}{2022}]{young2022}
{Young} A.~K.,  {Alexander} R.,  {Rosotti} G.,   {Pinte} C.,  2022, \mn@doi
  [\mnras] {10.1093/mnras/stac840}, \href
  {https://ui.adsabs.harvard.edu/abs/2022MNRAS.513..487Y} {513, 487}

\bibitem[\protect\citeauthoryear{{Zhu}}{{Zhu}}{2019}]{zhu2019}
{Zhu} Z.,  2019, \mn@doi [\mnras] {10.1093/mnras/sty3358}, \href
  {https://ui.adsabs.harvard.edu/abs/2019MNRAS.483.4221Z} {483, 4221}

\bibitem[\protect\citeauthoryear{{Z{\'u}{\~n}iga-Fern{\'a}ndez}
  et~al.,}{{Z{\'u}{\~n}iga-Fern{\'a}ndez} et~al.}{2021}]{zuniga2021}
{Z{\'u}{\~n}iga-Fern{\'a}ndez} S.,  et~al., 2021, \mn@doi [\aap]
  {10.1051/0004-6361/202141985}, \href
  {https://ui.adsabs.harvard.edu/abs/2021A&A...655A..15Z} {655, A15}

\makeatother
\end{thebibliography}




\bsp	
\label{lastpage}
\end{document}